%% file: main.tex
\documentclass[fleqn,usenatbib]{mnras}

\usepackage{newtxtext,newtxmath}

\usepackage[T1]{fontenc}
\usepackage{ae,aecompl}

\usepackage{graphicx}	
\usepackage{amsmath}	
\usepackage{gensymb}
\usepackage{color}
\usepackage{multirow}
\usepackage{hhline}
\defcitealias{Krumholz17}{KTOM}

\def\ga{\mathrel{\hbox{\rlap{\hbox{\lower4pt\hbox{$\sim$}}}\hbox{$>$}}}}
\def\la{\mathrel{\hbox{\rlap{\hbox{\lower4pt\hbox{$\sim$}}}\hbox{$<$}}}}

\usepackage[normalem]{ulem}

\definecolor{darkgreen}{rgb}{0.13, 0.55, 0.13}
\definecolor{orange}{rgb}{0.8, 0.15, 0.13}

\newcommand{\aref}[1]{\hyperref[#1]{Appendix~\autoref{#1}}}

\usepackage{scalerel,tikz}
\usetikzlibrary{svg.path}
\definecolor{orcidlogocol}{HTML}{A6CE39}
\tikzset{orcidlogo/.pic={
 \fill[orcidlogocol] svg{M256,128c0,70.7-57.3,128-128,128C57.3,256,0,198.7,0,128C0,57.3,57.3,0,128,0C198.7,0,256,57.3,256,128z};
 \fill[white] svg{M86.3,186.2H70.9V79.1h15.4v48.4V186.2z}
 svg{M108.9,79.1h41.6c39.6,0,57,28.3,57,53.6c0,27.5-21.5,53.6-56.8,53.6h-41.8V79.1z M124.3,172.4h24.5c34.9,0,42.9-26.5,42.9-39.7c0-21.5-13.7-39.7-43.7-39.7h-23.7V172.4z}
 svg{M88.7,56.8c0,5.5-4.5,10.1-10.1,10.1c-5.6,0-10.1-4.6-10.1-10.1c0-5.6,4.5-10.1,10.1-10.1C84.2,46.7,88.7,51.3,88.7,56.8z};
}}
\newcommand\orcidicon[1]{\href{https://orcid.org/#1}{\mbox{\scalerel*{
\begin{tikzpicture}[yscale=-1,transform shape]
\pic{orcidlogo};
\end{tikzpicture}
}{|}}}}

\title[Observable properties of winds]{The observable properties of cool winds from galaxies, AGN, and star clusters - II. 3D models for the multiphase wind of M82}

\author[Yuan et al.]{
Yuxuan Yuan$^{1}$~\orcidicon{0000-0001-6816-0682},\thanks{E-mail: yuxuan@mso.anu.edu.au (YY)}
Mark R. Krumholz$^{1,2}$~\orcidicon{0000-0003-3893-854X},
Crystal L. Martin$^{3}$
\\
$^{1}$Research School of Astronomy and Astrophysics, Australian National University, Canberra, ACT 2601, Australia\\
$^{2}$ARC Centre of Excellence for Astronomy in Three Dimensions (ASTRO-3D), Canberra, ACT 2601, Australia\\
$^{3}$Department of Physics, University of California, Santa Barbara, Santa Barbara, CA 93106, USA\\
}

\date{Accepted XXX. Received YYY; in original form ZZZ}

\pubyear{2021}

\begin{document}
\label{firstpage}
\pagerange{\pageref{firstpage}--\pageref{lastpage}}
\maketitle

\begin{abstract}

Galactic winds are a crucial player in galaxy formation and evolution, but observations of them have proven extraordinarily difficult to interpret, leaving large uncertainties even in basic quantities such as mass outflow rates. Part of this uncertainty arises from the relatively simplistic models to which complex wind observations are often fit, which inevitably discard much of the available information. Here we present an analysis of the wind of the nearby dwarf starburst galaxy M82 using a semi-analytic model that is able to take advantage of the full three-dimensional information present in position-position-velocity data cubes measured in the H~\textsc{i} 21 cm line, the CO $J=2\to 1$ line, and the H$\alpha$ line. Our best-fitting model produces position-dependent spectra in good agreement with the observations, and shows that the total wind mass flux in the atomic and molecular phases is $\approx 10$ M$_\odot$ yr$^{-1}$ (corresponding to a mass loading factor of $\approx 2-3$), with less than a factor of two uncertainty; the mass flux in the warm ionised phase is more poorly constrained, and may be comparable to or smaller than this. At least over the few kpc off the plane for which we trace the outflow, it appears to be a wind escaping the galaxy, rather than a fountain that falls back. Our fits require that clouds of cool gas entrained into the wind expand only modestly, suggesting they are magnetically confined. Finally, we demonstrate that attempts to model the wind using simplifying assumptions such as instantaneous acceleration and a constant terminal wind speed can yield significantly erroneous results.
\end{abstract}




\section{Introduction}
\label{cha:intro}

In 1963, Lynds and Sandage conducted the first H$\alpha$ observations of the central region of the starburst galaxy M82 \citep{Lynds&Sandage63} and discovered that the gas above and below the nucleus has been accelerated to a velocity much larger than the local escape speed. This paper marked the discovery of galactic winds, and shows that galaxies are not ``isolated islands'' as had been thought in the eighteenth century \citep{Wright1750}. Modern observations reveal that galactic winds are ubiquitous across nearly all starburst galaxies and active galactic nuclei (AGN), at both low and high redshift \citep{Veilleux05, Veilleux20, Heckman&Thompson17}, and these winds typically contain multiple phases within them. The hot phase, characterised by high temperature ($\sim 10^8$ K) and speed, is widely observed in X-ray emission, and carries most of the energy budget in galaxies with strong outflows. Cool phase outflows ($\sim 10 - 10^4$ K), on the other hand, are much denser than the hot component and carry most of the mass.

Galactic winds have a significant impact on the life-cycle of galaxies. They can expel gas and quench star formation in the galactic nuclei within tens of million years, making them a candidate to explain galaxies' low star formation efficiency and baryon fractions \citep{Hopkins11, Alatalo11}. They carry mass, metals and dust from the ISM  to the circumgalactic medium (CGM) or even intergalactic medium (IGM; \citealt{Nelson18b, Tumlinson11, Muratov17}), making the IGM the largest reservoir of baryons and metals across the universe \citep{Oppenheimer06}. This provides a possible solution to the long-standing ``missing baryons problem''. However, some of the mass entrained into winds may also form a galactic fountain, recycling back to galaxies and triggering further star formation \citep{Bregman80, Melioli15, Marasco17, Werk19, Li_M20}. 

In the first widely accepted galactic wind model, proposed by \citet{CC85}, the explosion energy produced by supernovae (SNe) drives a hot, adiabatically-expanding, fast outflow. This model describes the hot phase, but cannot explain the cool winds that are widely observed. Theoretical models for these began to appear several decades later, and there is still significant uncertainty about the origin of the cool phase. In one scenario, cool gas in the outflow launching region is rapidly shredded and shock heated into the hot winds, but this initial hot outflow then undergoes radiative cooling beyond a cooling radius of several kpc, resulting in the observed cool phase \citep{Thompson16, Schneider18}. A second possibility is that magnetic fields \citep[e.g.,][]{McCourt15, Banda-Barrag16, Banda-Barragan18a}, radiative cooling \citep[e.g.,][]{Scannapieco&Bruggen15, Gronke&Oh18, Gronke&Oh20b, Schneider20, Kim20}, or a combination of both \citep[e.g.,][]{Cottle20, Li_Z20, Kim20b, Banda-Barragan21a} allow cold gas to survive being shocked and entrained by a hot outflow. Yet a third scenario is that cool gas clouds are accelerated via radiation pressure \citep[e.g.,][]{Krumholz&Thompson13, Coker13a, Thompson15, Crocker18a} or cosmic ray pressure \citep[e.g.][]{Mao18a, Bruggen20a, Crocker21a, Crocker21b} from the central source, and are therefore accelerated gently and not shocked at all. In the latter two scenarios cool gas clouds maintain their phase structure along their trajectories, while in the first they disappear and re-form later.

Although numerical simulations provide the most accurate descriptions of galactic winds, they can survey at most a very limited portion of parameter space. This makes them unsuited to the task of extracting information from the observations that are ``sampled'' from a potentially much larger portion of parameter space. Hence analysis of the kinematic and thermodynamic structure of observed winds requires using analytic or semi-analytic models, which are capable of generating synthetic observations for comparison to observations with high enough computational efficiency to allow parameter fitting. However, these kinds of models often resort to highly simplified or heuristic prescriptions to minimize computational cost. For example, many models \citep[e.g.,][]{Steidel10, Prochaska11, Scarlata&Panagia15, Lochhaas18} describe winds as fully-filled spherically symmetric expanding shells, and adopt density, velocity and covering fraction prescriptions that are simple functions of radius from the center. \citet{Carr18} adopt a more realistic biconical geometry, but still rely on a simplified kinematic structure inherited from \citet{Scarlata&Panagia15}. Although these models are able to roughly reproduce stacked, spatially-unresolved spectra, they inevitably discard much of the information hidden within the spatially-resolved spectra available for nearby sources. Moreover, these models solely focus on a single phase of galactic wind, and hence cannot produce a unified picture of a multi-phase outflow. Prediction of the spatial and velocity structure of a multi-phase wind with enough complexity to allow useful model fitting, but with enough simplicity for that fitting to be computationally tractable, remains a great challenge for modern observational studies of galactic winds. 

This challenge led \citet[hereafter \citetalias{Krumholz17}]{Krumholz17} to propose a semi-analytical model that gives a reasonable balance between physical complexity and numerical efficiency. The \citetalias{Krumholz17} model can generate spatially resolved synthetic spectra based on a physical model for wind acceleration that depends on a few physical parameters and prescriptions, suitable for comparison to the resolved observations of wind. Unlike earlier semi-analytic models, it allows a wide range of wind geometries and driving mechanisms, and can model emission a variety of phases. This model is implemented as an extension of the open-source code Derive the Energetics and SPectra of Optically Thick Interstellar Clouds (\textsc{despotic}; \citealt{Krumholz14}). 

In this paper we provide a first test of this model against real observations, by using it to constrain the physical properties of the cool phases of the outflows in the dwarf starburst galaxy M82. We aim to extract much more detailed information about the properties of the wind than would have been possible with previous methods, taking full advantage of the three-dimensional position-position-velocity data cubes in multiple tracers that are available for one of the most well-observed galactic winds. We derive basic parameters such as mass outflow rate and wind geometry for each of the phases independently, and highlight the relationships between them.

The structure of the paper is as follows: In \autoref{cha:methods}, we briefly review the \citetalias{Krumholz17} model, the data to which we will apply it, and the methods we use to fit to the observed data. We then present the results from measurements in \autoref{cha:results}. We next discuss our results in \autoref{cha:discussion} and finally summarise our conclusions in \autoref{cha:conc}.

\section{Methods}
\label{cha:methods}

We first introduce the observational data to which we will fit in \autoref{ch2_sec:M82}. In \autoref{ch2_sec:gw_model} we summarise the \citetalias{Krumholz17} model and explain how varying its parameters affects predicted spectra. We describe our fitting method and compare different combinations of model choices in \autoref{ch2_sec:obs_diag}.

\subsection{Observations of the M82 outflow}
\label{ch2_sec:M82}

Messier 82 (M82) hosts one of the most well-studied galactic winds. M82 is an edge-on (inclination angle $\sim 80 \degree$) starburst galaxy \citep{O'Connell&Mangano78, O'Connell95}, located only 3.6 Mpc away (1 arcsec $\sim$ 17.5 pc) \citep{Freedman94}, making it a promising candidate for the study of outflow around the minor axis. \textit{Hubble Space Telescope} (HST) imaging reveals that M82 hosts very young massive clusters, concentrated in its central 500 pc (30 arcsec) nucleus. M82 is characterised by its intense central starburst, triggered by a tidal interaction with its neighbour M81. 

The M82 wind has been observed across the spectrum, from a hot phase observed in soft X-rays ($\sim 10^7$ K) \citep{Watson84, Bregman95, Strickland97, Lopez20}, to warm ionised gas traced by H$\alpha$ ($\sim 10^4$ K) \citep{McKeith95, Martin98, Westmoquette09}, to a cool neutral phase seen in H~\textsc{i} 21cm ($\sim 5000$ K) \citep{Martini18} and a cold molecular phase traced by CO ($\lesssim 100$ K) \citep{Walter02, Leroy15}. Recent observations suggest that the latter two phases exist largely on the edge of a central hot outflow, with H$\alpha$ lying at the interface between the hot outflow and surrounding cold gas.

\begin{figure*}
  \centering
  \includegraphics[width=1.\linewidth]{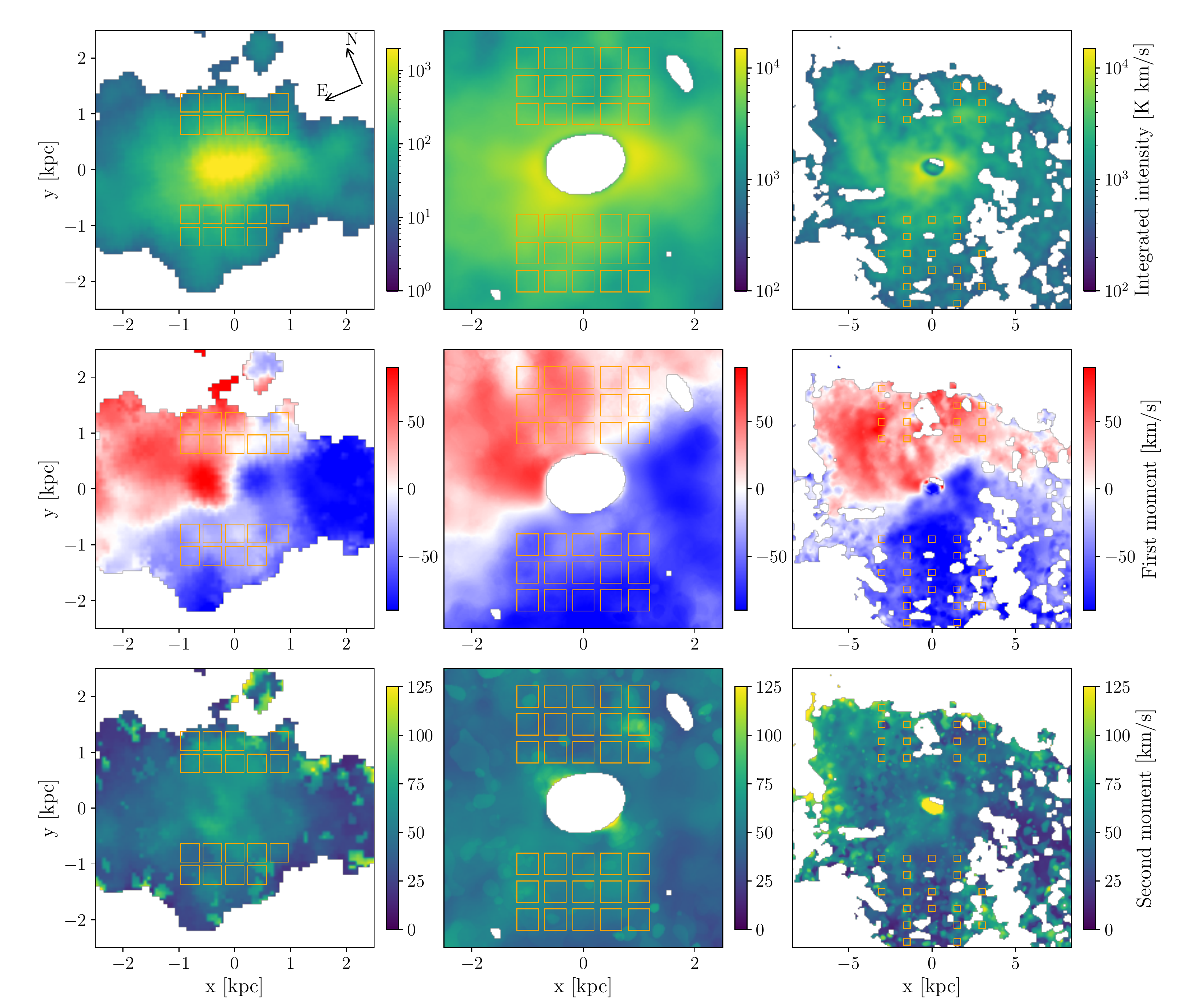}
\caption{Velocity-integrated  intensity  (top  row)  and first and second  moment  maps  (middle and bottom  row)  for  the  observations  of the CO $2\to1$ \citep[left]{Leroy15} and H~\textsc{i} 21 cm lines \citep[center and right]{Martini18}; the center and right columns both show the same data, and differ only in that the central column is zoomed in so that the field of view shown matches that for the CO in the left column, while the right column shows the full extent of the 21 cm data. We have oriented our coordinates so that the major and minor axis of the galaxy defined in \citet{Martini18} lie along the $x$ and $y$ axes, respectively; arrows in the top left panel show the orientation on the sky. We have also shifted the first moment so that a value of zero corresponds to the systemic velocity of 211 km s$^{-1}$. The white region in the centre of 21 cm map is masked because it is dominated by galaxy emission; other regions shown in white indicate non-detections. Orange squares indicate the apertures over which we compute average spectra for use in our fitting procedure; see main text for details.}
\label{ch2_fig:pos_spectra}
\end{figure*}

\subsubsection{Neutral phases: H~\textsc{i} and CO}
\label{ssec:neutral_phases}

In this paper we use the CO $J=2\to1$ data obtained by \citet{Leroy15} to trace the cold molecular phase of the outflow, and the H~\textsc{i} 21 cm data obtained by \citet{Martini18} to trace the cool neutral phase. The first data set consists of a position-position-velocity (PPV) cube that covers a $\approx 2.5\times 2.5$ kpc region around the galaxy and the second covers an $\approx 8\times 8$ kpc region. We refer readers to the original papers for full details on the observations. To show the overall kinematic structure of the cool wind, we plot the velocity-integrated intensity and first and second moment maps of these two data sets in \autoref{ch2_fig:pos_spectra}; we show two views of the H~\textsc{i} data, one zoomed in on the field that overlaps the CO data, and one zoomed out to show the full extent of the 21 cm data. We have oriented our coordinates so that the major and minor axes of the galaxy, as defined in \citet{Martini18}, lie along the $x$ and $y$ axes, respectively, and we set $v=0$ to correspond to the systemic velocity of 211 km s$^{-1}$. We can clearly see evidence for rotation both in the midplane and in the outflow region of the first moment map. We also see that the H~\textsc{i} outflow is much more extended than the molecular outflow, which suggests they have different spatial distributions and possibly different mass fluxes. 

We next select representative regions over which to compute averaged spectra to which we will compare our model. Our motivation for doing so is three-fold: first, averaging obviously increases the signal to noise ratio, yielding a cleaner fit. Second, while the semi-analytic model we use to generate predicted spectra as we vary our parameters is quite fast, it is not fast enough for it to be practical for us fit the full PPV cube at its native spatial resolution; we must reduce the size of the data for computational reasons. Third, while the \citetalias{Krumholz17} model we use is significantly more complex than previous galactic wind models, it is nonetheless obviously an oversimplification compared to reality. We therefore wish to pick regions that generally capture the rich information in the observation as a whole, but average out the data enough that our fitting is not overly biased by ``small scale'' spectroscopic features at certain positions, and gives a reasonable fit to the entire observation.

In accord with this principle, we extract spectra at a broad range of positions within two rectangular regions around the minor axis of the disc. We define two extraction regions; the first, which we apply to the $2.5\times 2.5$ kpc field within which the H~\textsc{i} and CO fields of view overlap, starts at projected distances 0.8 kpc off the disc ($|y|>0.8$ kpc); this geometric cut is to ensure that the emission to which we are fitting is dominated by the outflow and not the disc. The second extends from 2.5 kpc to 8 kpc, and covers the region within which we have only H~\textsc{i} data. In both regions, we stack spectra within square apertures with a size of $\pm 10''$ at each position, and we omit apertures where emission is undetected over any part of the aperture. We show the square regions over which we average in \autoref{ch2_fig:pos_spectra}, with the inner extraction region shown in the left and central columns, and the outer region in the right column. This procedure yields 30 (31) independent spectra for the H~\textsc{i} data set at central (full) region, and 18 for the CO (central region only).

\subsubsection{Warm ionised phase: H$\alpha$}

\begin{figure}
  \centering
  \includegraphics[width=\columnwidth]{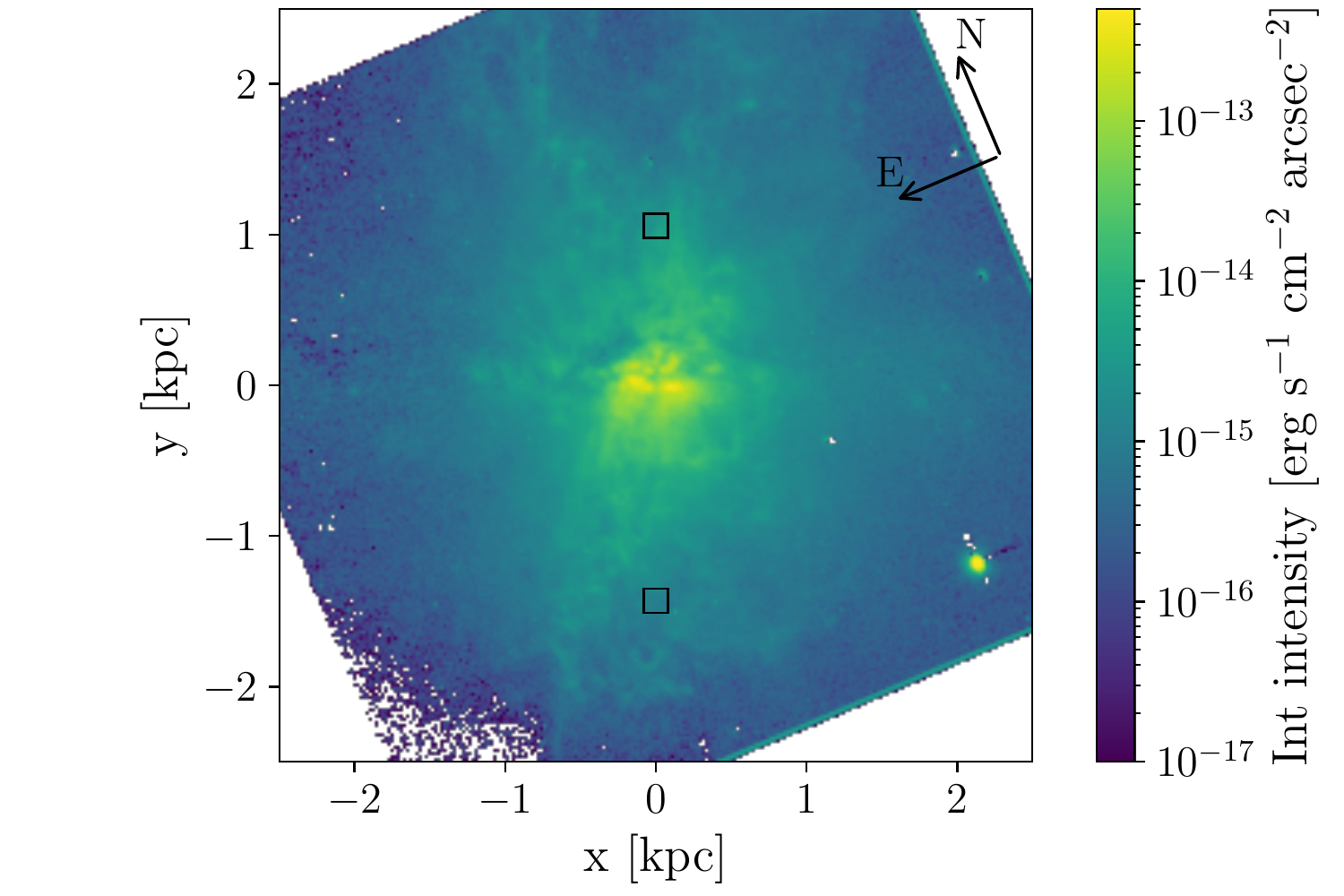}
\caption{Integrated intensity in the H$\alpha$ line \citep{Kennicutt08}. We have oriented the image to match the major and minor axes of the galaxy defined in \citet{Martini18}, the  $x$ and $y$ axes, respectively. Black squares indicate the apertures over which we compute average spectra for use in our fitting procedure \citep{Martin98}; see main text for details.}
\label{fig:Halpha}
\end{figure}

We used the M82-2 spectrum shown in Figure 1b of
 \citet{Martin98} to measure the kinematics of the warm ionised phase. The 
 3\farcm5 long slit was oriented along the outflow axis. Very high spectral resolution
 of 11.5 km/s was obtained using an echelle grating, and a narrowband filter blocked all
 orders except the one covering the H$\alpha$ line. This resolution reveals double peaked
 line profiles along the length of the slit, consistent with emission from the surface of an expanding cone (see \citep{Heckman90}). Figure 2d of \citet{Martin98} shows the velocity separation as a function of position
 along the slit. 
 
 We extracted a spectrum for each lobe at a location near the position of maximum velocity separation.  The black squares shown in Figure~\ref{fig:Halpha} mark these locations. These echellograms have an outdated format, so it was necessary to convert them to the
 FITS standard.\footnote{The code used to convert old IRAF images to FITS can be downloaded from https://github.com/mjfitzpatrick/imh2fits.} The spectra were not flux calibrated, so
 we scaled their intensity to the surface brightness measured in the same aperture of 
 H$\alpha$ image from \citet{Kennicutt08}.  


\subsection{Galactic wind line radiative transfer model}
\label{ch2_sec:gw_model}


We next summarise the most relevant features of the \citetalias{Krumholz17} wind model to which we will fit the data, leaving full details to the original paper and \citet{Thompson&Krumholz16}, from which it follows. For convenience we collect all of the parameters we introduce in this section in \autoref{tab:despotic_par}.

\begin{table*}
  \centering
  \input table/despotic_par.tex

  \caption{Summary of parameters for the \citetalias{Krumholz17} model. The top part of the table shows continuously-variable parameters, while the bottom shows discrete parameters. Full details on the definitions of the parameters are given in \autoref{ch2_sec:gw_model}. 
  }
  \label{tab:despotic_par}
\end{table*}

\subsubsection{Physical model}
\label{sssec:phys_model}

\citetalias{Krumholz17} describe a momentum-driven wind launched from a spherical region with enclosed mass $M_0$, radius $r_0$, and escape speed $v_0 = \sqrt{2 G M_0/r_0}$; the gas in this region has a mean surface density $\Sigma_0$, and is driven outward by a mechanism that injects momentum at a constant rate $\dot{p}$. The ratio of $\dot{p}$ to the momentum flux provided by gravity defines the generalised Eddington ratio
\begin{equation}
    \Gamma = \frac{\dot{p}}{4\pi G M_0 \Sigma_0}.
    \label{eq:gamma}
\end{equation}
The \citetalias{Krumholz17} model applies to a medium for which $\Gamma<1$, and thus a uniform medium would not be ejected into a wind. However, for a turbulent medium with Mach number $\mathcal{M}$, the surface density will be lognormally distributed, with a width that depends on $\mathcal{M}$, and material for which the surface density relative to the mean $x=\log \Sigma/\Sigma_0$ is sufficiently small will be driven outward. 

Mass that enters the wind begins at radius $r_0$ with velocity $v = 0$, but then accelerates under the action of the driving force. The resulting velocity evolves as
\begin{equation}
    2 u \frac{du}{da} = \frac{1}{a^2} \left(y \Gamma e^{-x} f_p - m\right),
    \label{eq:acclaw}
\end{equation}
where $u = v/v_0$, $a = r/r_0$, and $y$, $m$, and $f_p$ are functions that describe, respectively, how clouds of material entrained into the wind change their area as they move outward, how the gravitational potential varies with radius, and how the momentum deposition mechanism operates. We refer to the solution $u(a)$ to \autoref{eq:acclaw} as the wind acceleration law; \citetalias{Krumholz17} provide acceleration laws for several different choices of these functions, all of which we will consider here:
\begin{itemize}
    \item Cloud expansion: we consider $y = 1$, $y=a$, and $y=a^2$; the first of these possibilities corresponds to clouds that maintain constant area as they are accelerated, the third to clouds that maintain constant solid angle, and the second to an intermediate level of expansion.
    \item Gravitational potential: the acceleration law depends on the shape of the gravitational potential experienced by the wind, parameterised by the effective enclosed mass as a function of radius $m$ in \autoref{eq:acclaw}. \citetalias{Krumholz17} consider both $m=1$ (constant enclosed mass), corresponding to a point potential where gravity is dominated by mass in the acceleration region, and $m=a$ (enclosed mass increasingly linearly with radius), corresponding to an isothermal potential.
    \item Driving mechanism: we consider three ways a wind can be driven. The first is an idealised mechanism that simply deposits momentum at a fixed rate proportional to the cloud cross-sectional area; for this case $f_p=1$. The second is a wind driven by direct radiation pressure, in which case \citetalias{Krumholz17} show that $f_p = 1 - \exp(-e^{x}\tau_0/y)$, where $\tau_0 = \kappa \Sigma_0$ and $\kappa$ is the flux-mean specific opacity of the gas to radiation. The final possibility we consider is a cool wind driven by the momentum deposited by a hot outflow travelling at (dimensionless) speed $u_h$, for which $f_p = (1 - u/u_h)^2$.
\end{itemize}
Each of these choices affects the shape of the spectra produced by the wind. For example, as the cloud expansion law moves from constant area to constant solid angle, winds intercept more momentum from the central engine and thus accelerate more rapidly, pushing more emission to higher velocities; moving from a point to an isothermal potential, thereby decelerating gas more strongly via gravity, has the opposite effect. Different driving mechanisms lead to different shapes for line wings. We explore these differences in more detail below.

In addition to the acceleration law and the Mach number $\mathcal{M}$ of the medium from which the wind is launched, wind spectra are affected by two other primary factors: the wind mass flux and geometry. The mass flux $\dot{M}$ acts primarily to set the normalisation of the predicted spectra, although the relationship is somewhat more complex than this due to radiative transfer effects. The mass flux is determined by the surface density $\Sigma_0$ and by the value of $\Gamma$, which determines the minimum value of $x$ for which the right-hand side of \autoref{eq:acclaw} is positive. We refer to this value as $x_{\rm crit}$, and its value is given by $x_{\rm crit} = \ln\Gamma$ for ideal or hot gas-driven winds, and implicitly by the solution to $\Gamma e^{-x_{\rm crit}} [1 - \exp(-e^{x_{\rm crit}}\tau_0)] = 1$ for radiatively-driven winds. Following \citetalias{Krumholz17}, we choose to eliminate the dependence of $\dot{M}$ on $\Sigma_0$ by instead writing the wind mass flux in terms of the star formation rate $\dot{M}_*$, as
\begin{equation}
    \dot{M} = f_A \frac{\zeta_M}{\epsilon_{\rm ff}} \dot{M}_*,
    \label{eq:Mdot}
\end{equation}
where $f_A$ is the fraction of the area over which the wind can escape (i.e., where it is not blocked by dense material such as the remainder of M82's disc -- see below),
\begin{equation}
    \zeta_M = \frac{1}{2}\left[1-\mbox{erf}\left(\frac{-2x_{\rm crit}+\sigma_x^2}{2\sqrt{2}\sigma_x}\right)\right]
    \label{eq:zetaM}
\end{equation}
is the fraction of mass for which the right-hand side of \autoref{eq:acclaw} is positive, $\sigma_x$ is the width of the density distribution (determined by $\mathcal{M}$ -- see \citetalias{Krumholz17}, equations 11 and 12), and $\epsilon_{\rm ff}$ is the star formation rate per free-fall time. We adopt $\epsilon_{\rm ff}=0.01$ in this work, based on extensive observations showing that all star-forming systems fall near this value (see the review by \citealt{Krumholz19a} and references therein).

The final physical parameter for the wind is its geometry. Observed galactic winds are often biconical \citep[e.g.,][]{Shopbell&Bland-Hawthorn98}, with the outflow blocked from escaping the galaxy in the disc plane, and blowing out in directions normal to the plane. Moreover, cooler phases are often inferred to lie on the edge of a central region dominated by hot gas. For this reason, we consider winds with a ``cone sheath'' geometry, whereby the wind is confined between two cones with outer opening angle $\theta_{\rm out}\in (0,\pi/2]$ and inner opening angle $\theta_{\rm in}\in [0,\theta_{\rm out})$; note that we allow $\theta_{\rm in} = 0$ and $\theta_{\rm out}=\pi/2$, so we allow the possibility for a wind that is a fully filled cone or even a sphere covering a full $4\pi$ sr. The inner and outer opening angle determine the fraction of the unit sphere subtended by the wind: $f_A = \cos\theta_{\rm in} - \cos\theta_{\rm out}$. The central axis of the wind cone is inclined at an angle $\phi \in (-\pi/2,\pi/2)$ relative to the plane of the sky, where $\phi=0$ corresponds to the central axis of the outflow cone lying exactly in the plane of the sky. The values of these parameters also manifest in the spectral shape: if $\theta_{\rm in}$ is close to $\theta_{\rm out}$, this implies that the wind is confined to a narrow sheath, which manifests as spectra that are more narrowly peaked, since the line of sight passes through a smaller range of radii. Similarly, if $\phi = 0$ then the near and far sides of the wind are symmetric relative to the line of sight, implying symmetry in the red and blue sides of the spectrum; tipping the wind cone, $\phi\neq 0$, produces an asymmetry between the radius at which our line of sight passes through the near and far sides of the wind cone, which in turn induces an asymmetry between the red and blue sides of the spectrum.

Thus the physical model of the wind is parameterised by a total of three dimensional quantities -- $r_0$, $v_0$, and $\dot{M}_*$ -- and three dimensionless functions -- $y$, $m$, and $f_p$ -- and five or six dimensionless numbers $\Gamma$, $\mathcal{M}$, $\theta_{\rm out}$, $\theta_{\rm in}$, $\phi$, and, depending on the choice of driving mechanism, either $\tau_0$ or $u_h$. Most of these we will fit, as described below, but some can be fixed by other observations. We take $r_0 = 250$ pc, the observed radius of the star-bursting centre of M82 \citep{Kennicutt98}, and the escape speed from this region to be $v_0 = 170$ km s$^{-1}$ \citep{Greco12}, corresponding to a dynamical mass $M_0 = 8.2\times 10^8$ M$_\odot$. M82's star formation rate is $\dot{M}_* = 4.1$ M$_\odot$ yr$^{-1}$, which we derive by correcting the value inferred by \citet{Kennicutt98} from their assumed \citet{Salpeter55a} IMF to a \citet{Chabrier05} IMF.

In principle the shape of the gravitational potential is also measurable; however, in practice this has proven difficult. \citet{Sofue92a} find that the gas in the central 2 kpc of M82 is well-fit by Keplerian rotation, implying a point-like potential, while \citet{Greco12} find that the stellar rotation curve suggests a nearly-isothermal potential at radii $\lesssim 1$ kpc, giving way a point-like potential (constant enclosed mass -- see their Figure 5) beyond $\approx 2-4$ kpc. The differences between the gas and stellar velocity profiles are likely a result of the differential response of these components to either the tidal perturbation from M81 or the starburst itself. Moreover, both of these measurements are in-plane, and may have limited applicability to the out-of-plane potential, which is what matters for our purposes. The equipotential surfaces are almost certainly flattened, so that, to the extent that the potential follows \citeauthor{Greco12}'s model of a transition from isothermal to point-like at $\approx 2-4$ kpc in the plane, this transition should occur closer to the galactic centre in the vertical direction; however, since we do not know how flattened the potential is, we do not know how much closer. Consequently, we leave the potential shape as an unknown parameter to be fit.

\subsubsection{Chemical and excitation model}
\label{sssec:chemical}

In addition to the physical model, we require a model for the abundance and excitation state of the atoms and molecules that produce the observable emission. We have three different emission tracers -- H~\textsc{i}, CO, and H$\alpha$ -- which we discuss in turn.

We assume that H~\textsc{i} emission comes from material where all the hydrogen is atomic, and thus the mass per emitting atom is $\mu_{\rm H~\textsc{i}} = 2.3\times 10^{-24}$ g, as expected for the standard cosmic ratio of H to He. Given the low Einstein $A$ of the 21 cm transition, and the fact that H~\textsc{i} is efficiently thermalised by the Wouthuysen-Field effect \citep{Wouthuysen52a, Field59a}, we assume that the emitting atoms are in local thermodynamic equilibrium (LTE). We adopt a temperature of 5000 K, though this choice has no practical effect as long as the temperature is much larger than the $\ll 1$ K temperature difference between the two states.  Finally, the \citetalias{Krumholz17} model makes a distinction between ``correlated'' and ``uncorrelated'' winds; the former corresponds to cases where the each of the clouds entrained into the wind is individually optically thick (even if the wind as a whole might be optically thin because its filling factor is small), while the the latter to cases where the individual wind clouds are optically thin (even if the wind as a whole might be optically thick due to the superposition of many clouds along the line of sight). One can analogise the correlated case to a wind made up of opaque dust grains, and the uncorrelated case to a wind made up of transparent water droplets. Given the typically very low optical depth of warm H~\textsc{i}, we adopt the uncorrelated case for the 21 cm emission.

For CO we adopt a mass per CO molecule $\mu_{\rm CO} = 2.1\times 10^{-20}$ g, corresponding to a CO abundance of $1.1\times 10^{-4}$ CO molecules per H atom. We also assume LTE and adopt a temperature of $T=50$ K. These choices are somewhat more uncertain and significant than the equivalent ones for H~\textsc{i}, and we discuss their implications and the uncertainties they induce below. We also assume that CO is in the correlated case, given the typically high optical depth of CO-emitting clouds. It is worth noting that the \citetalias{Krumholz17} model does \textit{not} assume a fixed value of $\alpha_{\rm CO}$; instead, the value of $\alpha_{\rm CO}$ is computed self-consistently from the wind acceleration law using the large velocity gradient approximation.

Finally we come to H$\alpha$, which is the most complex case because it can be produced in two distinct phases. One is in fully ionised gas, and the other is in predominantly neutral gas that is subject to an ionising flux, and recombines at the ionised-neutral interface. The \citetalias{Krumholz17} model is intended to treat primarily the former, for this component we have mass per H$^+$ ion $\mu_{\rm H^+}=2.3\times 10^{-24}$ g, which gives an energy emission rate per unit volume $\Lambda n_{\rm H}^2$, where $n_{\rm H}$ is the number density of H nuclei, and the coefficient $\Lambda = 3.9\times 10^{-25}$ erg cm$^3$ s$^{-1}$. We further assume that the H$\alpha$ line is optically thin. Because the emission rate varies as the square of density, calculation of the H$\alpha$ emission requires that we adopt a clumping factor $c_\rho = \langle n_{\rm H}^2\rangle / \langle n_{\rm H}\rangle^2$, where the angle brackets indicate a volume average.\footnote{In the optical literature it is more common to define the filling factor $f = \langle n\rangle^2/\langle n^2\rangle = 1/c_\rho$. The two are obviously equivalent.} The emissivity is proportional to the value of $c_{\rho}$. We discuss our treatment of $c_{\rho}$ below.

For the second component, the emission will be proportional to the ionising flux to which the neutral gas is subjected. We have no direct knowledge of the ionising radiation field present in the wind of M82, so we make the simplest possible assumption that it is constant, in which case the emissivity at a given position and velocity is simply proportional to atomic hydrogen mass, and thus to the H~\textsc{i} 21 cm signal, in that voxel. The scaling between this signal and the emitted H$\alpha$ is unknown, and thus becomes an additional parameter to be fix.

Given these choices, we can compute a model-predicted spectrum of the wind in H~\textsc{i} 21 cm and CO $J=2\to 1$ emission at an arbitrary position using equations 56 and 78 of \citetalias{Krumholz17}, which are implemented in \textsc{despotic} \citep{Krumholz14}. We use the same procedure for H$\alpha$, except that we add the emission predicted by \textsc{despotic} to an emission term that is proportional to the 21 cm signal at the corresponding position and velocity.


\subsection{Fitting method}
\label{ch2_sec:obs_diag}

We next outline our strategy for fitting the model to the data. We consider a series of wind models described by all possible combinations of driving mechanism, potential and expansion law. For each combination, we optimise the set of physical parameters under a Bayesian framework. We then identify the combination that gives the best fit between model and data. We carry out this exercise independently for the Northern and Southern sides of the wind, since the wind is observed to be asymmetric between these two hemispheres \citep{Leroy15, Martini18}.

\subsubsection{Parameter fitting}
\label{ch2_ssec:mcmc_method}

For the cases of H~\textsc{i} and CO, we fit five parameters, $\phi$, $\theta_{\mathrm{in}}$, $\theta_{\mathrm{out}}$, $\mathcal{M}$, and $\Dot{M}$, for ideal momentum driven winds, with an additional parameter $\tau_0$ for radiation driven and $u_h$ for hot gas driven winds, respectively; note that we do not need to treat $\Gamma$ or $x_{\rm crit}$ as independent parameters, because they can be deduced from the other parameters via \autoref{eq:Mdot} and \autoref{eq:zetaM}. The case of H$\alpha$ is handled slightly differently, and we defer a discussion of it to \autoref{sssec:Ha_method}. We adopt uniform priors for $\sin(\phi)$ ($\phi \in (-\pi/2, \pi/2)$), $\theta_{\mathrm{in}} \in [0, \pi/2)$, $\theta_{\mathrm{out}} \in (\theta_{\rm in}, \pi/2)$, $\log\mathcal{M} \in (0,4)$\footnote{In principle we could adopt an informative prior for $\mathcal{M}$, since the velocity dispersions and approximate temperatures of the various gas phases are measureable quantities. However, we shall see below that all other quantities are essentially uncorrelated with $\mathcal{M}$, so there is little point to doing so.}, and $\log \Dot{M}/{\rm M}_\odot\,{\rm yr}^{-1} \in (-1, 3)$. We note that the range of mass outflow rate is broad enough to encompass essentially all previously reported results for mass outflow rates \citep[e.g.,][]{Muratov15, Fluetsch19, Roberts-Borsani19, Roberts-Borsani20, Roberts-Borsani20a}.
We also adopt flat priors on $\tau_0$ subject to the constraint $\tau_0 > 1/\Gamma$ (since a wind is only launched at all if this condition is met), and flat priors on $u_h \in (1,50)$, corresponding to hot wind speeds from $170 - 8,500$ km s$^{-1}$.

In addition to these model parameters, we allow the zero of velocity to vary with position. The reason we do so is that there is considerable evidence that the outflow from M82 is rotating \citep{Leroy15}, whereas our \citetalias{Krumholz17} model does not include rotation. We approximate that the effect of rotation is simply to shift the zero point of the spectrum as a function of position, and include this effect by leaving the zero point at each position as a free parameter to be optimised.

We compute the likelihood of model given the observations on a spectrum-by-spectrum basis, so the total likelihood function is given by the product of the likelihoods over all sky positions and velocity bins,
\begin{equation}
\label{eq:likelihood}
\mathcal{L}(f_\mathrm{obs}| \mathbf{Q}) = -\frac{1}{2} \sum_{i} \sum_{j} \frac{ \left[ f_{i}(v_{j} - v_{0,i}; \mathbf{Q})-f_{i,\mathrm{obs}}(v_{j}) \right]^2 }{ \sigma_{ij}^2 }
\end{equation}
where $\mathbf{Q}$ is the vector of parameters being fit ($\phi$, $\theta_{\rm in}$, $\theta_{\rm out}$, $\mathcal{M}$, $\dot{M}$, $\tau_0$ or $u_{\rm h}$ for radiatively-driven or hot gas-driven models), $f_{i}$ and $f_{i,\rm obs}$ represent the predicted and observed spectra at the $i$th position (i.e., at the $i$th box shown in \autoref{ch2_fig:pos_spectra}), the spectra are measured at velocities $v_j$, and $\sigma_{ij}$ is the observational uncertainty at position $i$ and velocity bin $j$. At each position $i$, we evaluate the sum using the velocity shift $v_{0,i}$ that maximises the likelihood at that position. The posterior PDF can then be calculated via Bayes' theorem
\begin{equation}
P ( \mathbf{Q} |f_{\mathrm{obs}}) \propto
\mathcal{L}(f_{\mathrm{obs}} | \mathbf{Q}) P_{\mathrm{prior}}(\mathbf{Q})
\end{equation}{}

We calculate the posterior probability distribution function, from which we determine the best-fitting parameters with their uncertainties, using the affine-invariant Markov Chain Monte Carlo (MCMC) ensemble sampler \textsc{emcee} \citep{Foreman-Mackey13}. We repeat this calculation for each possible combination of driving mechanism, potential and expansion law, using 40 MCMC walkers, each iterating over 500 steps to sample the posterior distribution of parameters. In general we find that the distribution of walkers needs $\sim$ 50 iterations to stabilise, so we burn in the first 100 iterations and derive the posterior PDFs from the rest.

\subsubsection{Which combination of driving mechanism, potential and expansion law?}
\label{ch2_ssec:dist_comb}

Having optimised physical parameters for all combinations of the driving mechanism, potential, and expansion law, denoted as (dm, p, ex), we are now in a position to distinguish which combination gives the best fit to the observed data. In order to decide this, for each combination (dm, p, ex), we find the largest value of the likelihood function returned by any of the MCMC sample points, denoted as $\hat{\mathcal{L}}_{\mathrm{(dm, p, ex)}}$, and compute the corresponding Akaike information criteria (AIC) (see \citealt{Sharma17})
\begin{equation}
\mathrm{AIC}_{\mathrm{(dm, p, ex)}} = 2k - 2\ln\hat{\mathcal{L}}_{\mathrm{(dm, p, ex)}}
\end{equation}
where $k=5,$ 6 and 6 are the number of parameters for ideal, radiation and hot gas driven winds, respectively: 3 parameters to describe the cone sheath geometry, 1 parameter for the Mach number, 1 parameter to describe the mass-outflow rate, and 1 extra parameter to describe the optical depth or hot gas velocity for the radiation-driven or hot gas cases, respectively. The corresponding Akaike weight for each combination then is
\begin{equation}
\begin{aligned}
w(\mathrm{dm, p, ex}) &= \frac{\mathrm{e}^{-\Delta_{\mathrm{(dm, p, ex)}}/2}}{\sum\limits_{\rm (dm, p, ex)} \mathrm{e}^{-\Delta_{\rm (dm, p, ex)} / 2}}   \\
\Delta_{\mathrm{ (dm, p, ex)}} &= \mathrm{AIC}_{\mathrm{(dm, p, ex)}}-\min\left(\mathrm{AIC}_{\mathrm{(dm, p, ex)}}\right)
\end{aligned}
\end{equation}
which gives the relative probability for each combination of driving mechanism, potential, and expansion law. 

\subsubsection{Modifications for H$\alpha$}
\label{sssec:Ha_method}

As discussed in \autoref{sssec:chemical}, H$\alpha$ is substantially more complex than H~\textsc{i} or CO, both because it potentially arises from two distinct phases -- fully ionised gas and the surfaces of neutral clouds -- and because the former of these has an emissivity that scales as the square of the local volume density, and thus involves a clumping factor $c_\rho$. Our method for handling the emission from H~\textsc{i} cloud surfaces is straightforward; as discussed in \autoref{sssec:chemical}, we assume that this emission is proportional to the H~\textsc{i} mass at every PPV voxel, and the constant of proportionality between the 21 cm signal and the H$\alpha$ signal simply becomes another parameter to be fit by the MCMC.

The clumping factor $c_\rho$ requires more care; since it simply applies an overall scaling factor to the signal, it is essentially degenerate with the mass outflow rate. For this reason, if we were to treat $c_\rho$ as a fit parameter, our derived mass outflow rate would depend sensitively on our prior for $c_\rho$. For this reason, we choose to carry separate fits using a range of fixed $c_\rho$ values that span the plausible range, and allow us to determine a corresponding plausible range of mass outflow rates. We discuss the results of this experiment below.

\section{Results}
\label{cha:results}

In this section, we fit our model to each of the observational data sets described in \autoref{ch2_sec:M82}, using the fitting method described in \autoref{ch2_sec:obs_diag}. We examine the H~\textsc{i} data first in \autoref{ch3_sec:wn_ph}, since they are the most straightforward conceptually, then move on to the CO data in \autoref{ch3_sec:mol_ph} and the H$\alpha$ data in \autoref{ch3_sec:wi_ph}.

\begin{table*}
  \centering
  \input table/res_fitting_HI.tex

  \caption{Best-fitting parameters and Akaike weights obtained for various combinations of driving mechanism (Column 1), potential (Column 2) and expansion law (Column 3) for the warm neutral phase outflow. The top part of the table shows results for the Northern hemisphere; here we show all possible combinations of driving mechanism, potential, and expansion law, and highlight the cases with the highest Akaike weights $w$ (Column 4). The bottom part of the table is for the Southern hemisphere, and here we omit models with $w<0.01$, which are ruled out with $>99\%$ confidence, for brevity. The fit quantities shown in Columns 5 - 11 are the inclination angle $\phi$ (Column 5), inner and outer opening angle of the wind $\theta_{\rm in}$ and $\theta_{\rm out}$ (Columns 6 and 7), logarithm of mass outflow rate $\log\dot{M}$ measured in $M_\odot$ yr$^{-1}$ (Column 8), logarithm of Mach number $\mathcal{M}$ (Column 9), wind optical depth $\tau_0$ (Column 10; radiatively-driven models only), and hot gas dimensionless velocity $u_h$ (Column 11; hot gas-driven models only). For most fit quantities, values are specified with the 50th percentile value given first, and the 84th to 50th and 16th to 50th percentile ranges listed as the superscript and subscript values, respectively; thus for example the entry $4.73^{+4.21}_{-5.10}$ for $\phi$ in the (Ideal, Point, Area) case indicates a 50th percentile value $\phi = 4.73^\circ$, and a 16th to 84th percentile range $\phi = -0.37^\circ$ to $8.94^\circ$. The exception is $\mathcal{M}$, where in most cases our posterior probability distribution is bounded above only by the prior we impose; for these cases we report only a 16th percentile lower limit.}
  \label{tab:res_fit_hi}
\end{table*}

\begin{figure}
  \centering
  \includegraphics[width=\columnwidth]{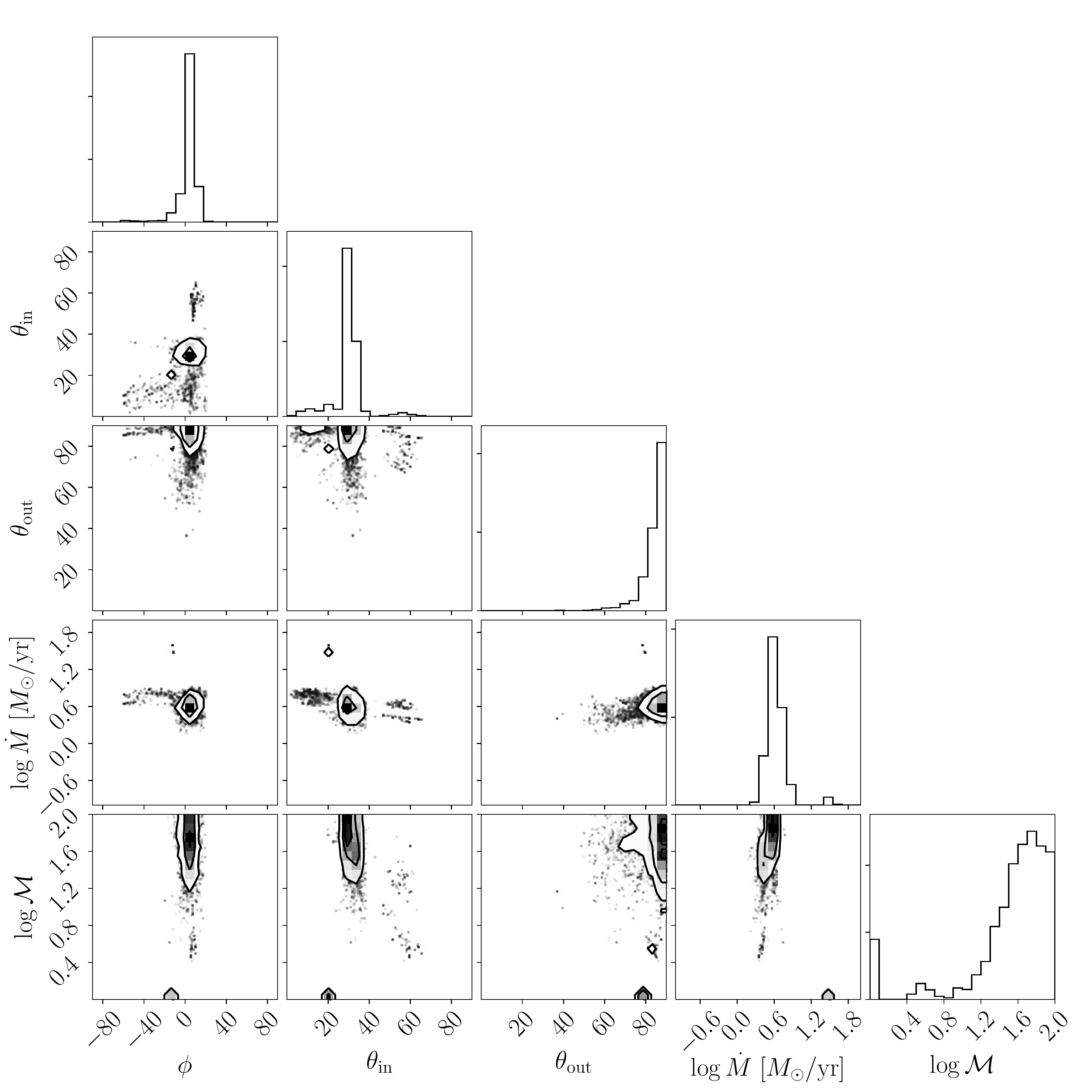}
\caption{Corner plot showing the 1D and 2D histograms of the posterior PDFs of all parameters (see \autoref{tab:despotic_par}) describing the Northern side of the warm neutral phase outflow traced by H~\textsc{i} emission. The model fit shown is for an ideal wind in a point gravitational potential, for clouds that maintain constant area as they flow outward; this combination of driving, potential, and cloud expansion yields the highest Akaike weight as compared to other combinations. }
\label{fig:corner_hi}
\end{figure}

\subsection{Warm neutral phase}
\label{ch3_sec:wn_ph}

As discussed in \autoref{ssec:neutral_phases}, for the H~\textsc{i} data we define four distinct regions: the Northern and Southern Hemispheres of the wind, and for both an inner region that overlaps with the CO and H~$\alpha$ data, and an outer region where only H~\textsc{i} is detected. We fit each region independently. For simplicity we first discuss our results for the Northern inner region, using this region as an example to explain and explore our method in \autoref{sssec:HI_north_inner} - \autoref{ch3_ssec:hi_dm_p_ex}.
We then present results for the the inner Southern region in \autoref{sssec:hi_south}, and for the outer extraction region where only H~\textsc{i} is detected in \autoref{sssec:hi_large_scale}. Since we are fitting the two hemispheres separately, we divide the values of $\dot{M}$ returned by our MCMC fits by a factor of two, since we are fitting only half the available solid angle at a time; all values of $\dot{M}$ listed below include this factor of two division.

\subsubsection{Fitting results}
\label{sssec:HI_north_inner}

We summarise the best-fit parameters, uncertainties, and Akaike weights returned by our pipeline as applied to the central region of the Northern side of the outflow in \autoref{tab:res_fit_hi}. As discussed in \autoref{ch2_ssec:mcmc_method}, we fit each possible combination of driving mechanism ($f_p$), gravitational potential ($m$), and expansion law ($y$) separately, so we report a best fit for each combination. However, from the Akaike weights shown in the table, we find that by far the best fit to the data comes from a point gravitational potential and clouds that maintain constant area; the Akaike weights corresponding to the three possible driving mechanisms (ideal, radiative, and hot gas) are similar enough that our fits do not allow us to distinguish between them. We show the joint and marginal posterior PDFs for our fit parameters for the ideal, point potential, constant area case in \autoref{fig:corner_hi}; the results for the radiatively-driven and hot gas-driven cases are qualitatively identical. We see that the fit parameters are well constrained and largely non-degenerate, with most of them converged into a tiny island of parameter space, with the exception of $\mathcal{M}$, for which we only obtain a lower limit that $\log\mathcal{M} \gtrsim 1.4$.

Our best-fitting geometric parameter values are largely consistent with and extend the results of previous observational studies. The near-zero orientation angle $\phi$ indicates the edge-on nature of the M82 and the large opening angles $\theta_{\rm in}$ and $\theta_{\rm out}$ implies the geometry of the outflow is approximately a conical sheath, as has been proposed before. The mass outflow rate at the nuclear region is about $4$ M$_\odot$ yr$^{-1}$, corresponding to a mass-loading factor of $\eta_{\rm wn} =  \Dot{M}/\Dot{M_*} \approx 1.0$, which is also within the range of values derived in previous literature \citep{Strickland&Heckman09}. However, our uncertainty range is far small than in previous work -- even taking the extreme ranges of all of our acceptable models, our fit requires $\dot{M}$ to lie between $3.0$ and $6.2$ M$_\odot$ yr$^{-1}$ with 68\% confidence; we have therefore constrained the mass outflow rate to $\approx 50\%$. We remind readers that this is the mass outflow rate for the atomic phase on the Northern side of the wind only; we return to other phases and to the Southern hemisphere below.

\subsubsection{Model validation}
\label{ch3_ssec:hi_model_valid}

\begin{figure*}
  \centering
  \includegraphics[width=\textwidth]{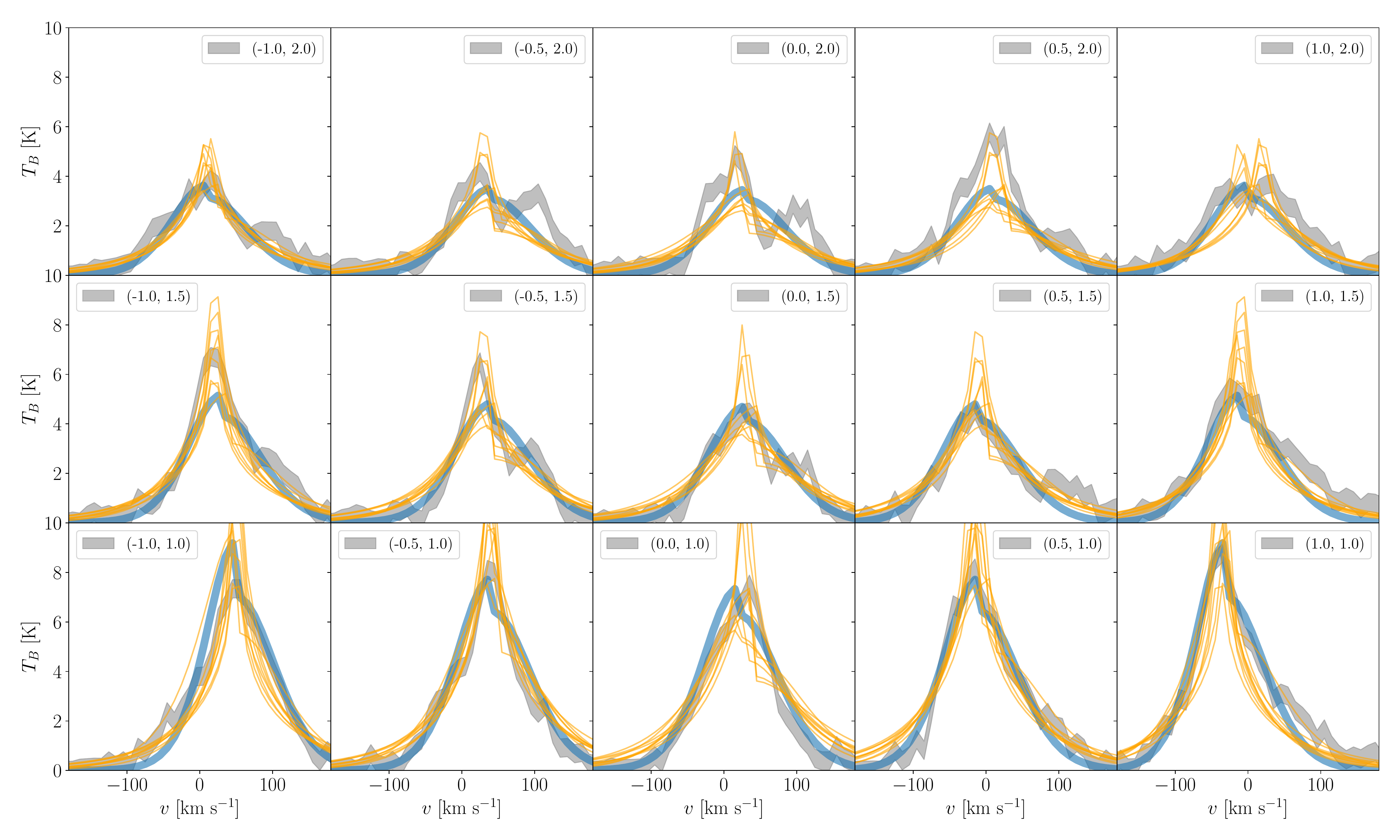}
\caption{Predicted versus observed H~\textsc{i} 21 cm spectra. Each panel shows the spectrum at one of our sample positions, expressed as brightness temperature $T_B$ as a function of velocity $v$, with $v=0$ corresponding to the systemic velocity of M82. The panels are arranged in the same pattern as in \autoref{ch2_fig:pos_spectra}, i.e., the upper left panel here is the spectrum at the position corresponding to the upper leftmost of the orange boxes indicated in \autoref{ch2_fig:pos_spectra}; legends in each panel give the $(x,y)$ coordinates of the box centre, in units of kpc. Blue lines show the predicted spectra for the set of parameters that gives the largest posterior probability found by our MCMC fit, while orange lines show spectra predicted using the parameters of 10 random walkers at the last iteration of MCMC sampling. For comparison, we also show the observed spectrum with its $1\sigma$ errors (grey band).}
\label{fig:spec_hi_comp}
\end{figure*}

Before accepting the best-fitting parameters we have obtained as definitive, it is important to establish that the our parametric model gives a reasonable description of that data. To this end, we compare our model-predicted spectra to the observed ones in \autoref{fig:spec_hi_comp}. In this figure, the grey band shows the observed spectrum (with $1\sigma$ errors) at each of our sample positions, while the blue and orange lines show predicted spectra computed using parameters from the single highest-likelihood model found by the MCMC (for the ideal, point potential, constant area case), and from 10 randomly-selected walkers at the final iteration of the MCMC, respectively. We see that the theoretical spectra give a fairly good fit to the observations at all positions, both in strength and shape.

\begin{figure*}
  \centering
  \includegraphics[width=\textwidth]{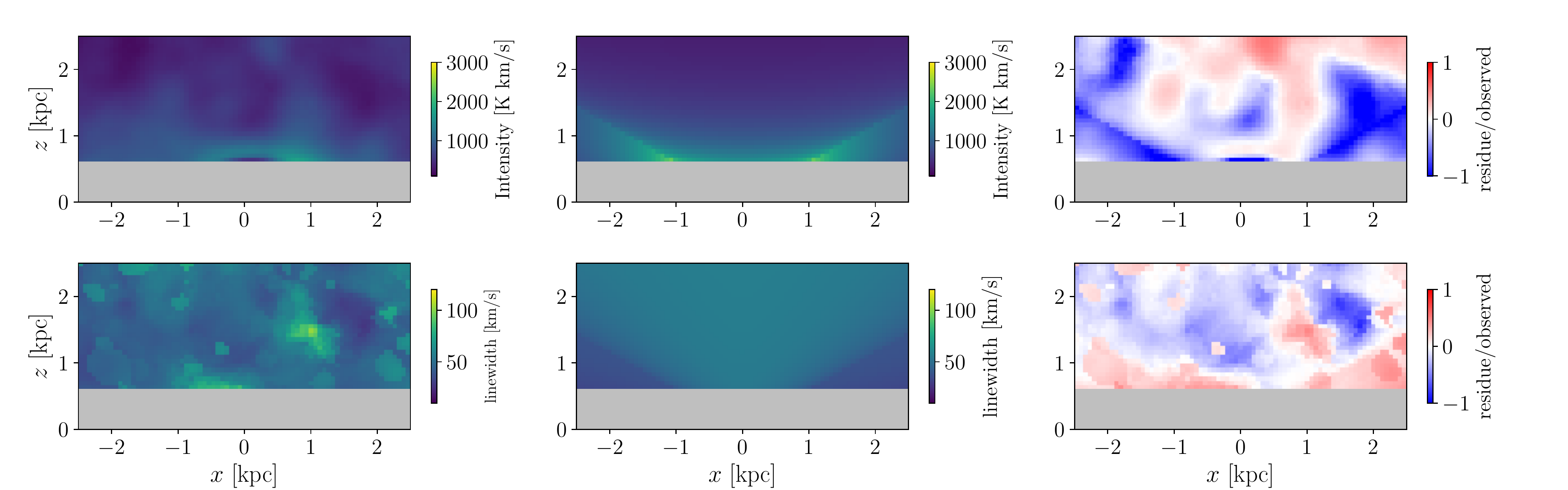}
\caption{Velocity-integrated brightness temperature (top row) and brightness temperature-weighted second moment map (bottom row) for the observations (left), theoretical predictions using our best-fitting model (middle) and the residual between the two normalised by the observations (right) for the Northern side H~\textsc{i} 21 cm line data. The black lines show our best-fitting angles $\theta_{\rm in}$ and $\theta_{\rm out}$ for the inner and outer angles of the outflow cone. The grey central region is a mask to block out $0-0.6$ kpc around the disc midplane, where disc emission is significant.}
\label{fig:moment_hi}
\end{figure*}

Having shown that our model can provide a reasonable fit to the spectra at our chosen positions, we next investigate whether our fits give a good match to the entire 2D map, or, in other words, whether our model can reproduce the data that are not included in the fits. In \autoref{fig:moment_hi} we compare the observed integrated intensity and intensity-weighted second moment maps to model maps generated using our highest-likelihood combination of parameters (blue line in \autoref{fig:spec_hi_comp}). In this figure, we mask the region $z=0 - 0.6$ kpc to avoid contamination from the disc. Outside the masked region, we see that the predicted and observed maps show reasonably similar morphology and absolute value, illustrating that our model successfully reproduces the bulk structure of the outflow.
We do see that our model predicts significantly more limb-brightening than is observed in the integrated-intensity map, and a sharper decrease in second moment as one moves away from the central axis. This may be due to our oversimplification of the true geometry, as we assume there is a hard cutoff of wind material at the inner and outer boundary. At a given height, this feature of our model implies increasing line of sight distance and decreasing velocity range as the impact parameter ($x$) approaches the inner cone, which overestimates the velocity-integrated intensity and underestimates the second moment. This artificial boundary effect is not present in real galaxies, as the geometry is more likely to resemble a biconical frustum (for which the ``base'' in the plane of the galaxy is a disc of finite area, rather than a point), and the boundary of the volume occupied by H~\textsc{i} is much less sharp than in our idealised model.

\begin{figure*}
  \centering
  \includegraphics[width=\textwidth]{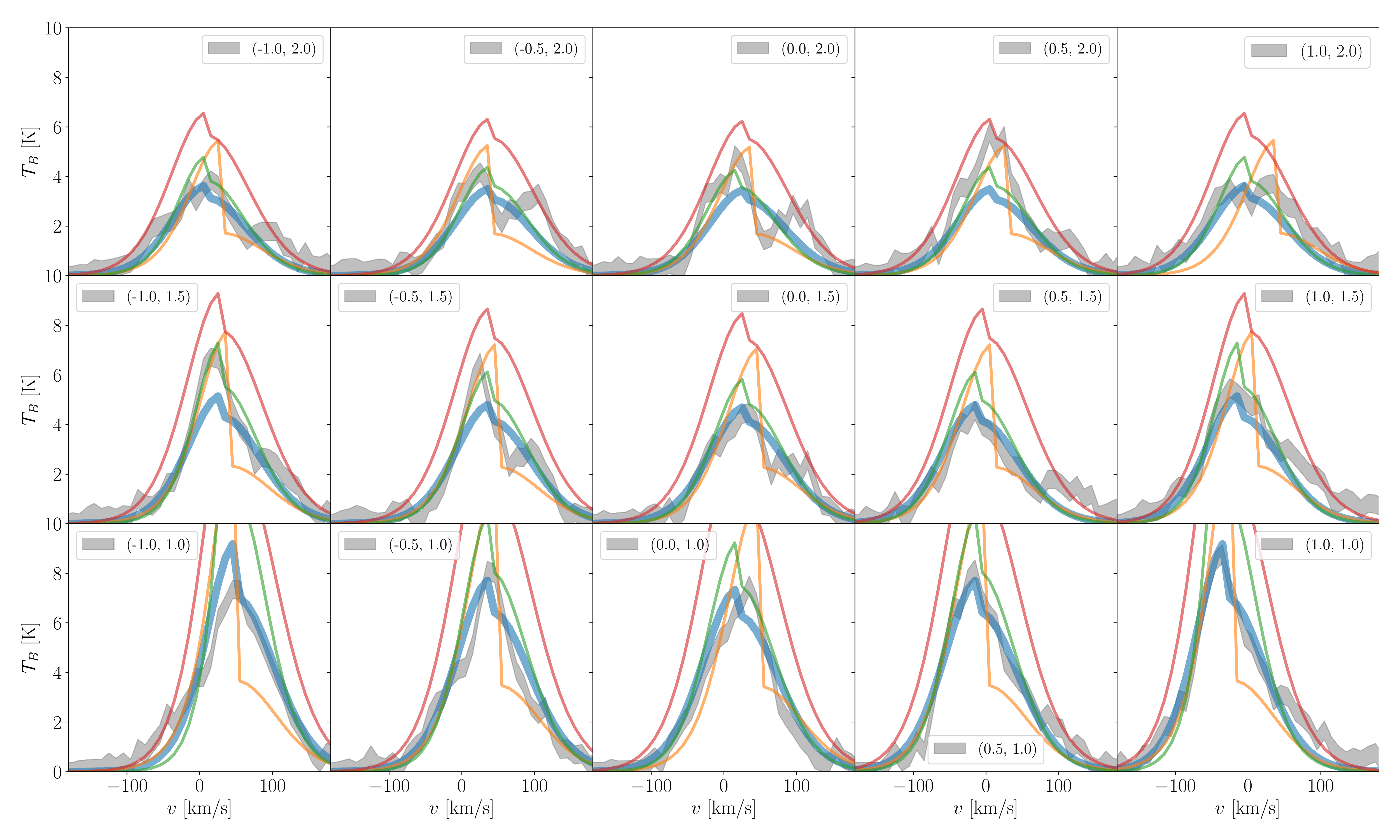}
\caption{Same as \autoref{fig:spec_hi_comp}, but now showing how the predicted spectra vary in response to changes in the fit parameters. The blue line and grey region are the same as in \autoref{fig:spec_hi_comp}, and show the highest-likelihood model and the observations, respectively. Other colours show predicted spectra produced by starting from the highest-likelihood parameters and increasing $\phi$ by $10^\circ$ (orange), decreasing $\theta_{\rm in}$ and $\theta_{\rm out}$ by $10^\circ$ (green), and doubling $\Dot{M}$ (red).}
\label{fig:spec_hi_par}
\end{figure*}

\subsubsection{Understanding parameter constraints}

We next examine how model-predicted spectra vary as we change parameters, with the goal of both justifying the tight constraints returned by our MCMC fits, and gaining insight into which features of the observed spectra provide constraints on particular parameters. In
\autoref{fig:spec_hi_par}, we show predicted spectra produced by taking the single highest likelihood model (blue line in \autoref{fig:spec_hi_comp}) and changing the parameters on which our model gives tight constraints one by one.
We first increase the inclination angle $\phi$ by 10 degrees, thereby tilting the outflow axis relative to the plane of the sky (orange lines in \autoref{fig:spec_hi_par}). We see that the increased value of $\phi$ gives rise to an asymmetry in the spectra. The reason for this is simple: positive velocities correspond to the far side of outflow while negative velocities correspond to the near side. A small value of $\phi$ means we are observing both near and far side of wind at similar distances from the nuclear region, so only optical depth effects, which are negligible in the case of H~\textsc{i}, could induce asymmetry between the negative and positive velocity sides of the spectra. However, for non-zero $\phi$, our line of sight passes through the the near and far sides of the outflow at two different distances from the nucleus. The near and far sides therefore have different emissivities, leading to a sharp change in the spectrum near zero velocity. Such a sharp feature is not observed, which is why our model prefers an orientation angle $\phi$ close to zero.

The green line in \autoref{fig:spec_hi_par} shows the effect of decreasing both the inner and outer opening angles $\theta_{\rm in}$ and $\theta_{\rm out}$ by 10$^\circ$. This change makes the profile narrower for lines of sight close to the central axis of the wind (central columns in \autoref{fig:spec_hi_par}), and both narrower and more intense near $v=0$ for lines of sight farther from the central axis (leftmost and rightmost columns). The physical origin of this behaviour is that, as we narrow the outflow cone at fixed $\dot{M}$, we both reduce the amount of material for which the velocity vector is well-aligned to the line of sight (thus depressing the high-velocity wings of the line) and create denser outflows, increasing the intensity overall, particularly along heavily limb-brightened lines of sight. Thus both the shape of the line wings and the amount of variation in intensity from one line of sight to another serve to constrain the opening angle. 

Finally, we see that, as expected, the mass outflow rate acts as primarily as a normalisation of the spectra, affecting the overall intensity while leaving the shape of the profile mostly unchanged.\footnote{The shapes of the spectra are not completely insensitive to $\dot{M}$ because higher $\dot{M}$ requires higher $\Gamma$, i.e., that the momentum flux be closer to the Eddington value; this in turn manifests as somewhat different wind acceleration laws. However, this is a second-order effect.} With increased $\Dot{M}$, the generated spectra overestimate the observed ones at nearly all velocities and positions. This strong effect explains why our model is able to produce such strong constraints on $\dot{M}$.

\begin{figure*}
  \centering
  \includegraphics[width=1.\linewidth]{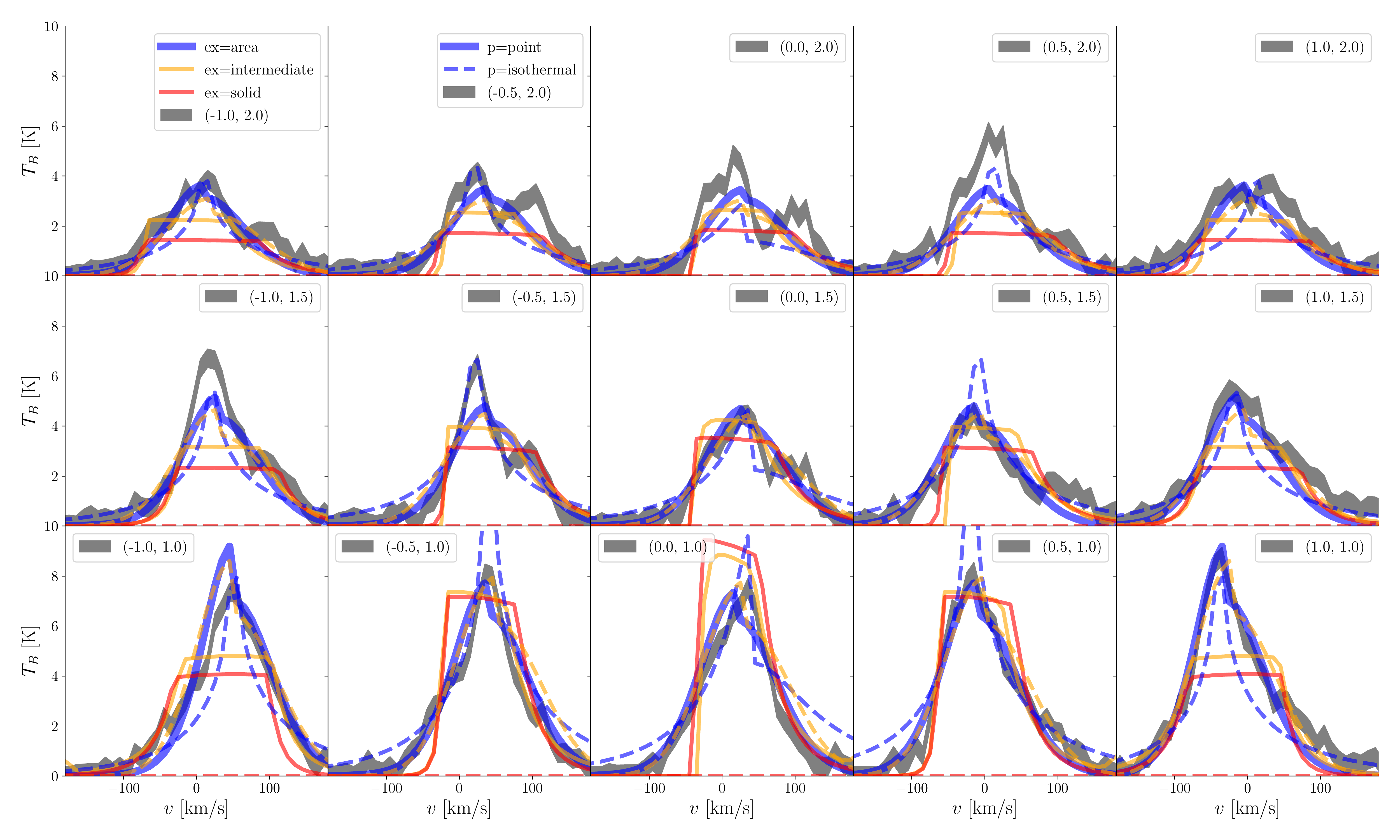}
\caption{Same as \autoref{fig:spec_hi_comp}, but now showing how the predicted spectra vary in response to the choices of potential and expansion law, for ideal winds. We show the best-fitting spectra for each individual combination of potential and expansion law. The spectrum produced with the combination that yields the highest Akaike weight is shown by the thickest line and the observed spectrum with 1$\sigma$ errors is shown by the grey region, as in \autoref{fig:spec_hi_comp}. Spectra adopting point and isothermal gravitational potentials are shown as solid and dashed lines, respectively, while those adopting constant area, intermediate, and constant solid angle expansion laws are shown in blue, yellow and red line, respectively.
}
\label{fig:spec_hi_p_ex}
\end{figure*}

\subsubsection{Understanding constraints on the potential and expansion law}
\label{ch3_ssec:hi_dm_p_ex}

We have seen that our fitting not only provides strong constraints on wind parameters for fixed combinations of driving mechanism ($f_p$), gravitational potential ($m$), and expansion law ($y$), comparison between the models clearly favours particular potentials and expansion laws. As a reminder, the former describes the shape of the gravitational potential, and thus the rate of gravitational deceleration, as the wind moves out. The latter dictates how clouds' cross-sectional areas, and thus the amount of momentum per unit time they intercept from the wind driving mechanism, change as they flow outward. 

To understand why we are able to constrain these parameters, it is helpful to begin by examining \autoref{tab:res_fit_hi}. In this table, we see that some models
yield unrealistically large orientation angles $\phi \sim \pm 40^\circ$ with a huge uncertainty, while others yield $\phi \approx 0$, consistent with the edge-on nature of M82. The reason for this is simple: we observe emission near zero velocity in all spectra, which implies either that there is some gas that truly has near-zero velocity at all positions, or that the zero-velocity emission is coming from gas whose velocity vector lies entirely in the plane of the sky. The first option is only possible for certain combinations of cloud expansion and potential: examining the acceleration law, \autoref{eq:acclaw}, we see that acceleration of the wind can remain close to zero at all radii only for the combinations $y=1$, $m=1$ (constant area, point potential), $y=1$, $m=a$ (constant area, isothermal potential), and $y=a$, $m=a$ (intermediate area, isothermal potential) -- and it is precisely these combinations that produce $\phi\approx 0$. For all other combinations, the presence of emission at zero velocity forces the MCMC fit to select models that have substantial amounts of material launched along the plane of the sky, which, for a conical sheath geometry, in turn requires $\phi \gg 0$ or $\phi \ll 0$. However, this compensation worsens the fit to the remainder of the spectrum, yielding a lower Akaike weight overall.

We illustrate this effect in \autoref{fig:spec_hi_p_ex}, which shows how the quality of fits varies with the different choices of potential and expansion law, for ideal winds. There are six lines, corresponding to each possible combination of potential (indicated by solid versus dashed) and expansion law (indicated by colour); each line corresponds to the set of parameters with the highest likelihood recovered by the MCMC for that combination of expansion law and potential. We see that although all models produce near zero-velocity emission, the models where the wind is more strongly accelerated ($y = a$ or $y=a^2$) clearly show a deficit of emission at the centre, and a line shape function completely different to that of the observed spectra. In contrast, the combination $y=1$, $m=1$ that has the highest Akaike weight provides a much better comparison to the observations, both at line centre and line wings. Our method is therefore able to distinguish between different combinations of cloud expansion and potential with confidence.

\subsubsection{Results for the Southern hemisphere}
\label{sssec:hi_south}

Thus far our discussion has been limited to the Northern side of the wind. We use the same method to analyse the Southern hemisphere, again limiting ourselves for now to the inner region that overlaps the CO data. We summarise the best-fit parameters, uncertainties, and Akaike weights for the Southern side of the outflow in \autoref{tab:res_fit_hi}. For brevity here we show results only for models with Akaike weight $w>0.01$, since all other models are ruled out with $>99\%$ confidence. As for Northern hemisphere, we find that the best fit to the data comes from a point gravitational potential and clouds that maintain constant area, and the Akaike weights corresponding to the first two possible driving mechanisms are still similar enough that our fits do not allow us to distinguish between them. However, unlike in the Northern hemisphere, the fit parameters for the South converge to more than one island, though within each island they are well constrained and largely non-degenerate. Overall we find that our model provides a somewhat poorer, less constrained fit to the Southern Hemisphere than to the North, likely because the wind in the South suffers greater tidal forces from M81 that produce structures not captured in our model; these tidal effects are evident in the data, which show significant asymmetry about the minor axis of M82, and a significant blue-shift compared to the Northern Hemisphere. However, for the largest island, we still recover near-zero orientation angles, and most importantly, mass outflow rates with values similar to those of Northern hemisphere, illustrating the robustness of out fit. We further note that there are changes to some parameters compared to the Northern Hemisphere, especially for the opening angles, which change by $\Delta \theta_{\rm in} ~ 10^\circ$ and $\Delta \theta_{\rm out} ~ -20^\circ$, which shows the asymmetric geometry of the wind system. 

\subsubsection{Results for the large-scale extraction region}
\label{sssec:hi_large_scale}

We have thus far presented our fits for the central region of M82 where data for all three phases overlap. However, warm neutral phase traced by H~\textsc{i} emission extends well beyond the region over which the other phases are detected. Therefore it is worthwhile to investigate this single phase outflow to the greatest extent we can achieve; in addition to providing further information by itself, this allows us to perform a useful consistency check, since our inner and outer extraction regions represent non-overlapping parts of the wind. This exercise therefore tests whether our results are reproducible when we look at a different part of the same structure. We therefore repeat our analysis for the Northern and Southern outer extraction regions.

\begin{figure}
  \centering
  \includegraphics[width=\columnwidth]{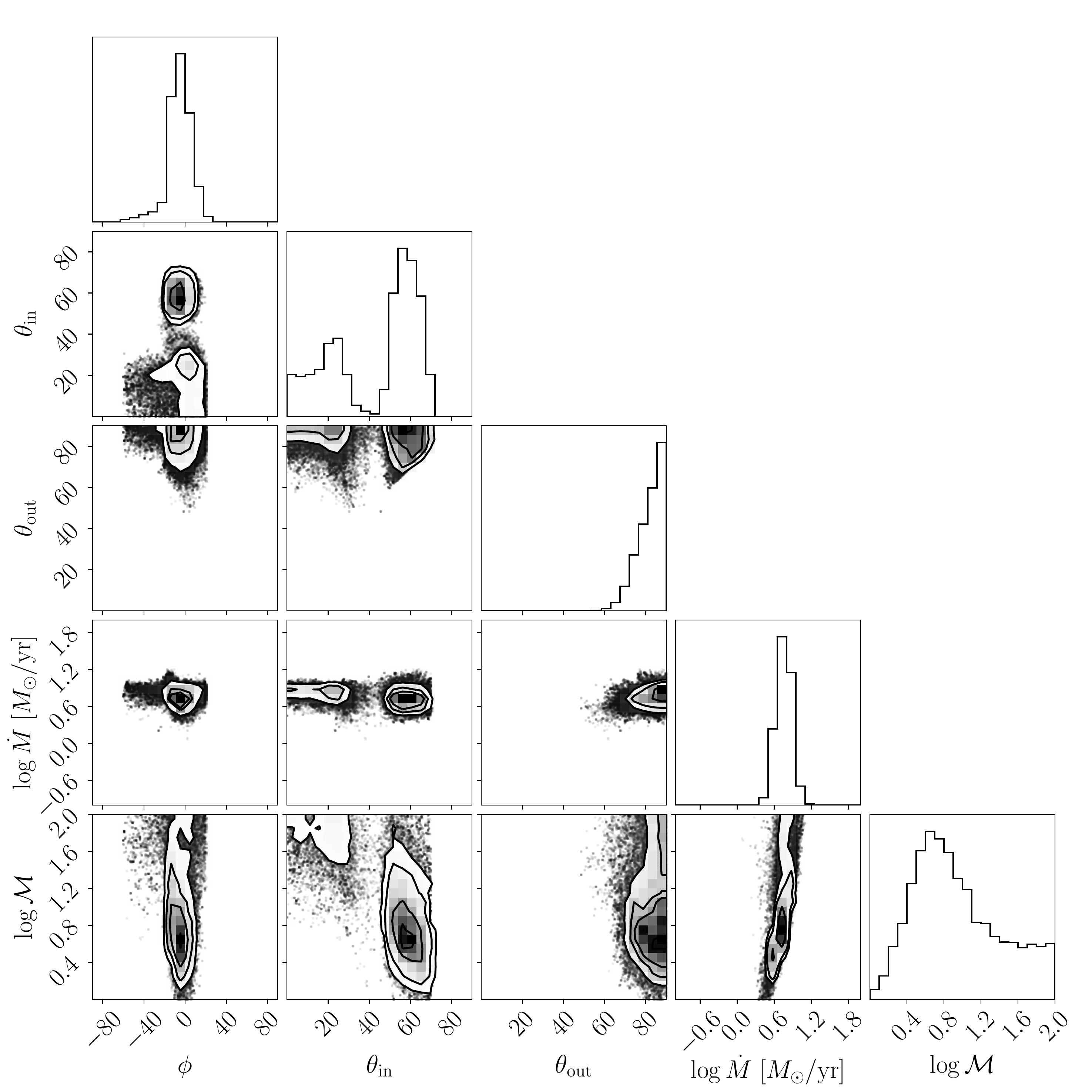}
\caption{Same as figure~\autoref{fig:corner_hi}, but show the results for H~\textsc{i} at larger scale}
\label{fig:corner_hi_large}
\end{figure}

As in \autoref{sssec:HI_north_inner}, we summarise the fitting results in \autoref{tab:res_fit_hi}; the table reports fits for the large-scale region on both the North and South sides, and we again omit models with Akaike weights $<0.01$ for brevity. From the models remaining, we again find that clouds of constant area being driven out of a point potential provide the best fit to the data (though models with intermediate expansion are not ruled out quite as strongly as for the inner region), and we are unable to differentiate strongly between different possible wind acceleration mechanisms. We show the joint and marginal posterior PDFs for our fit parameters for the ideal, point, area case for the Northern hemisphere in \autoref{fig:corner_hi_large}; results for the Southern hemisphere are qualitatively similar. We see that the fit parameters converge into two islands of parameter space characterised by a near-zero orientation angle $\phi$ and reasonable values of opening angles $\theta_{\rm in}$ and $\theta_{\rm out}$, verifying that the warm neutral phase of the outflow also displays a cone sheath geometry at large scale. The mass outflow rate is approximately $6.2$ M$_\odot$ yr$^{-1}$, corresponding to a mass-loading factor of $\approx 1.50$. This mass outflow rate is consistent at the $1\sigma$ level with our previous estimate based on the inner part of the wind, which demonstrates the robustness of our model to different scales we are fitting. The constraints on the geometry are also consistent at the $1\sigma$ level, but we find that our fits to the larger-scale data have much larger uncertainties, $\sim 20^\circ$ for $\theta_{\rm in}$ and $\sim 10^\circ$ for $\phi$ and $\theta_{\rm out}$. This may be caused by the stronger tidal torque induced by M81 on the larger scale.

\begin{figure*}
  \centering
  \includegraphics[width=1.0\textwidth]{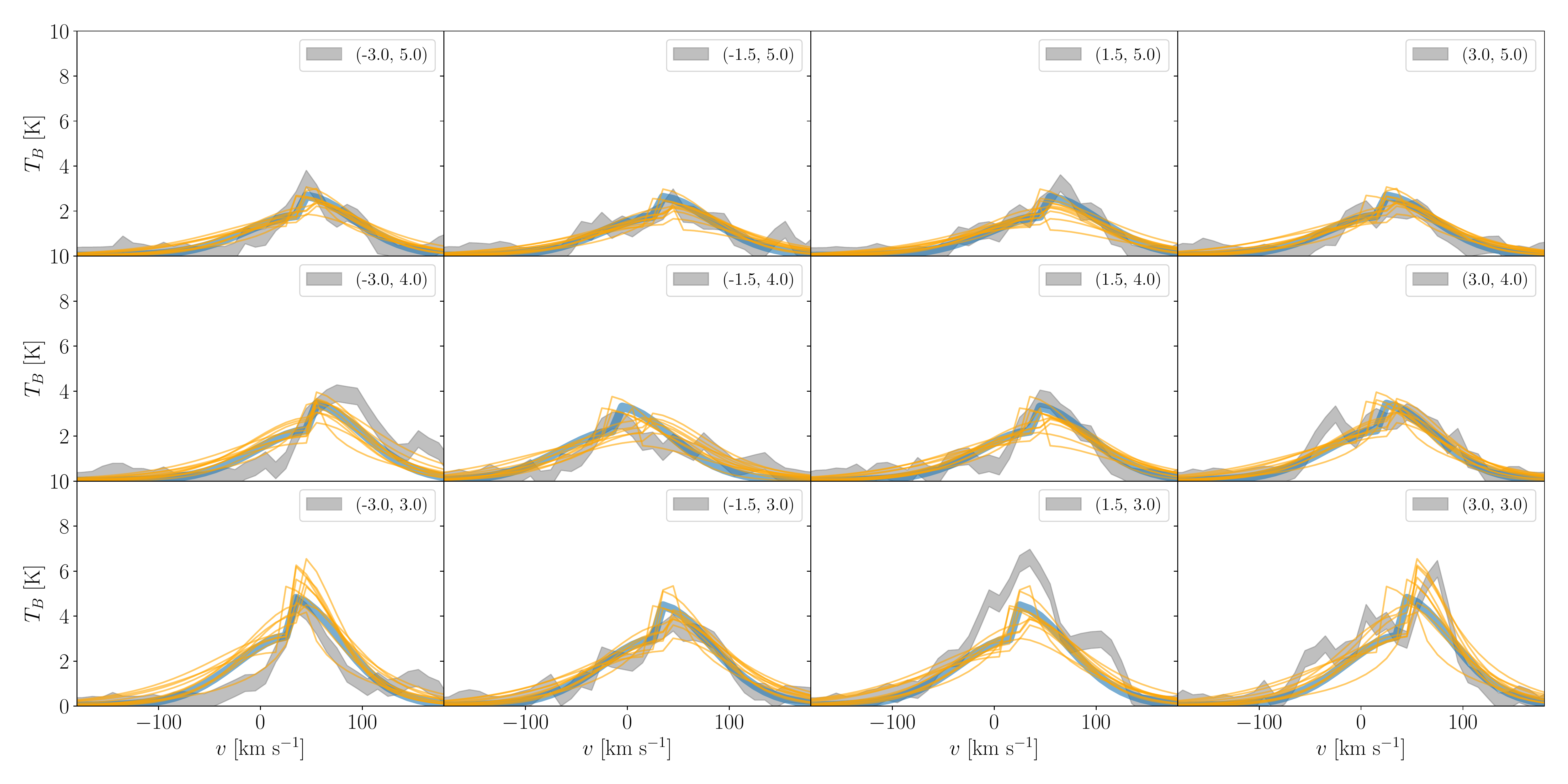}
\caption{Same as \autoref{fig:spec_hi_comp}, but showing the results for H~\textsc{i} at larger scale. As in \autoref{fig:spec_hi_comp}, each panel corresponds to one of the boxes shown in \autoref{ch2_fig:pos_spectra}.}
\label{fig:spec_hi_large}
\end{figure*}

In \autoref{fig:spec_hi_large} we compare predicted and observed spectra for the best-fitting case for the Northern hemisphere, just as we did in \autoref{fig:spec_hi_comp}. We again see that the theoretical spectra give a fairly good fit to the observations at all positions.

The relationship between the Northern and Southern hemispheres of the wind on the large scale are very similar to those on small scale. The Southern hemisphere gives comparable mass outflow rate and slightly different opening angle compared to the North. 

\subsection{Cold molecular phase}
\label{ch3_sec:mol_ph}

We next repeat the main steps in the analysis presented in \autoref{ch3_sec:wn_ph} for the cold molecular phase as traced by the CO $J=2\to1$ observations. 

\begin{table*}
  \centering
  \input table/res_fitting_CO.tex

  \caption{Same as table~\autoref{tab:res_fit_hi}, but show the results of fits for CO.}
  \label{tab:res_fit_co}
\end{table*}

\begin{figure}
  \centering
  \includegraphics[width=\columnwidth]{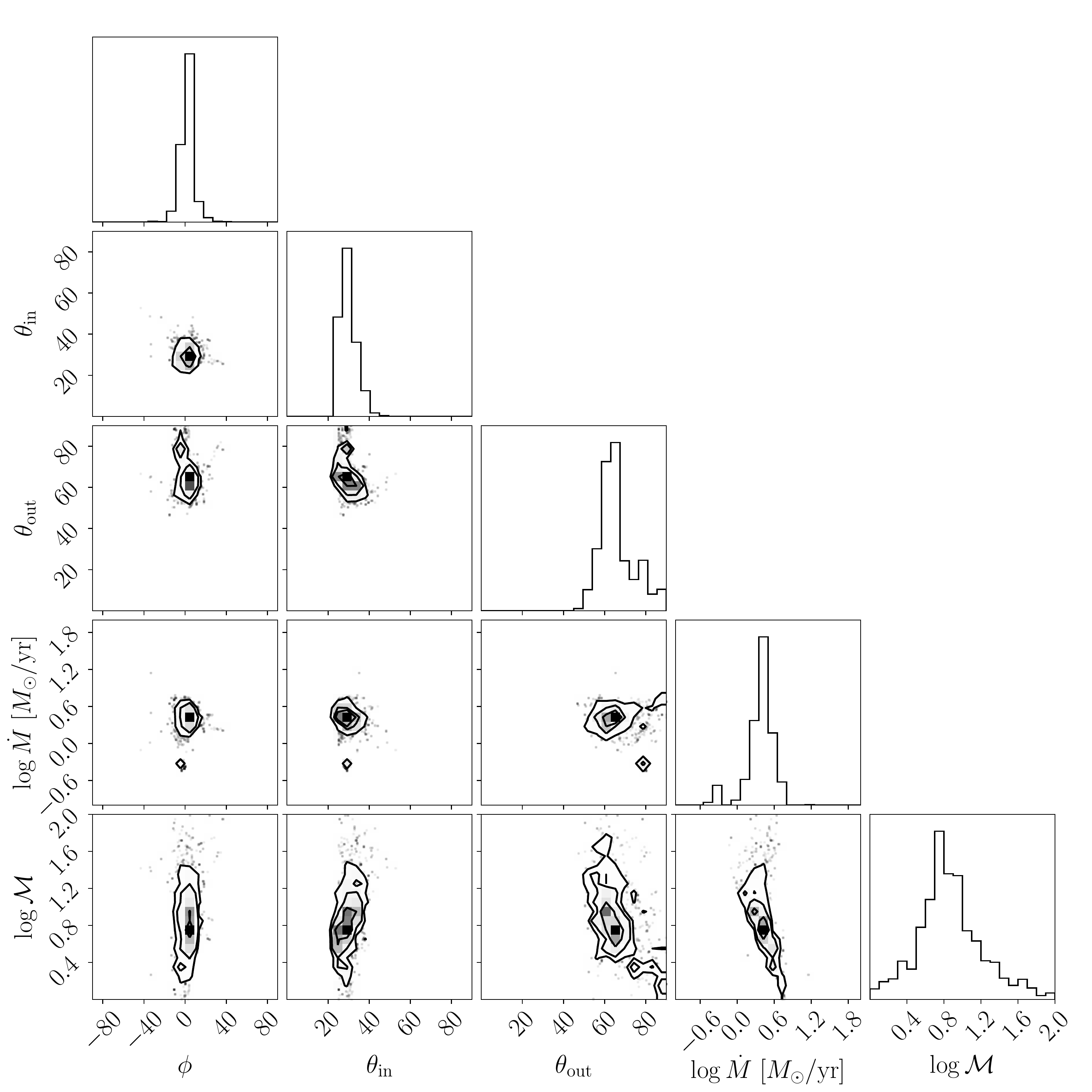}
\caption{Same as figure~\autoref{fig:corner_hi}, but show the results for CO.}
\label{fig:corner_co}
\end{figure}

Similar to \autoref{ch3_sec:wn_ph}, we summarise the fitting results to the Northern side of the outflow in \autoref{tab:res_fit_co}. We omit models with Akaike weights $<0.01$ for brevity, and from the models remaining, we again find that clouds of constant area being driven out of a point potential provide the best fit to the data, and that we are unable to differentiate strongly between different possible wind acceleration mechanisms. We show the joint and marginal posterior PDFs for our fit parameters for the ideal, point, area case in \autoref{fig:corner_co}. Again, we see that the fit parameters converge into a tiny island of parameter space characterised by a near-zero orientation angle $\phi$ and reasonable values of opening angles $\theta_{\rm in}$ and $\theta_{\rm out}$, verifying that the molecular phase of the outflow also displays a cone sheath geometry \citep{Walter17}. The mass outflow rate is approximately $2.2$ M$_\odot$ yr$^{-1}$, corresponding to a mass-loading factor of $\eta_{\rm cm} =  \Dot{M}/\Dot{M_*} \approx 0.55$. As with H~\textsc{i}, the mass outflow rate is very well-constrained; considering the extremes of all three models with acceptable Akaike weights, it lies between $1.3$ and $3.2$ M$_\odot$ yr$^{-1}$ with 68\% confidence.

\begin{figure*}
  \centering
  \includegraphics[width=\textwidth]{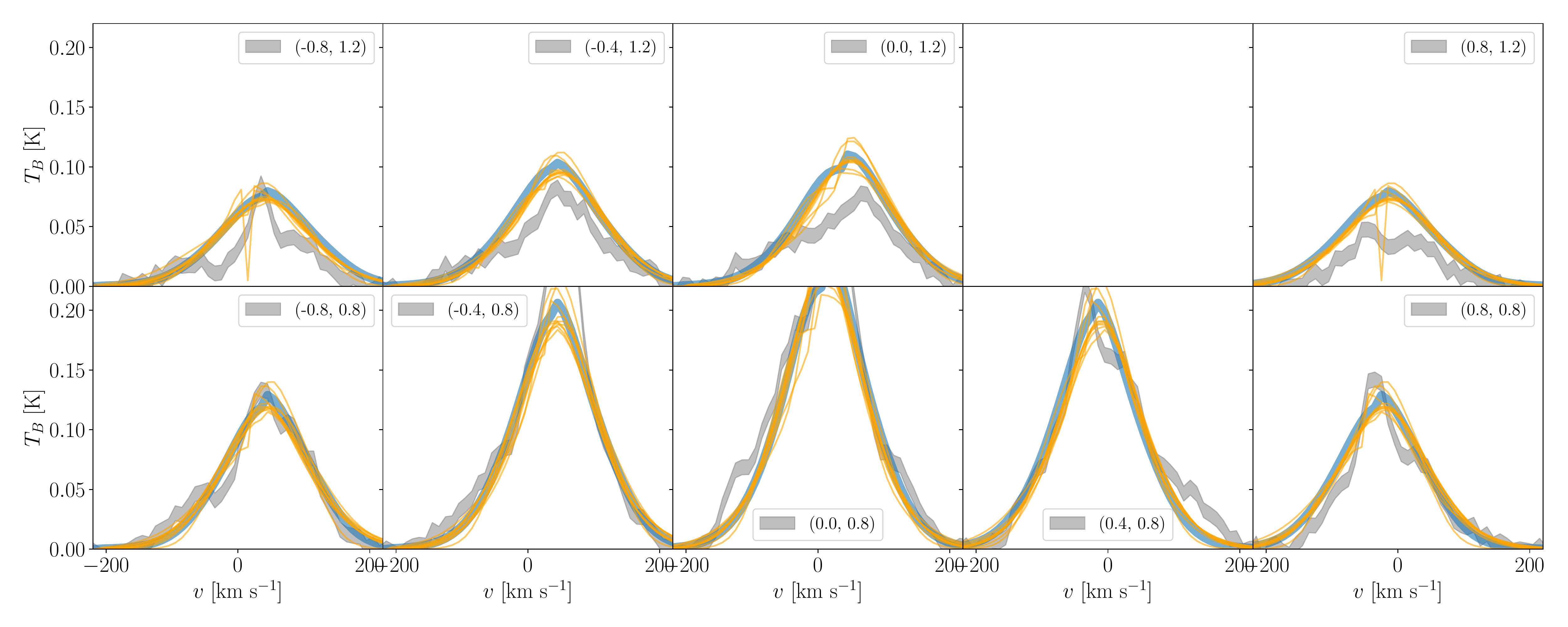}
\caption{Same as \autoref{fig:spec_hi_comp}, but showing the results for CO. As in \autoref{fig:spec_hi_comp}, each panel corresponds to one of the boxes shown in \autoref{ch2_fig:pos_spectra}.}
\label{fig:spec_co}
\end{figure*}

In \autoref{fig:spec_co} we 
compare predicted and observed spectra for the best fitting case, just as we did in \autoref{fig:spec_hi_comp} for the H~\textsc{i} data. We again see that the theoretical spectra give a fairly good fit to the observations at all positions, though the quality of the fits for H~\textsc{i} is somewhat better than that for CO. This is due to the greater complexity of modeling CO lines, which require us to make several approximations that may fail in certain parts of the wind. First, we rely on the large velocity gradient (LVG) approximation to evaluate optical depths, which becomes invalid near zero velocity, where thermal broadening becomes important. Second, we assume that the molecules are in LTE at a fixed temperature at all positions; both assumptions likely become invalid in low-density regions, where the collision rate is too low to thermalise the molecules, or to efficiently radiate away the energy deposited by shocks. The failure of this assumption may explain why our models fit some positions better than others. 

Finally, and most importantly, we caution that our need to choose a fixed temperature for the CO molecules likely induces a systematic uncertainty that exceeds the uncertainties quoted above. To first order, for an optically thick line such as CO $J=2\rightarrow 1$, we expect the emissivity per unit mass to scale close to linearly with the gas temperature, so a factor of two uncertainty on the gas temperature corresponds roughly to a factor of two uncertainty in the mass flux. Since the confidence intervals we obtain from our fits are considerably narrower than this, the gas temperature likely represents the single largest source of uncertainty remaining. Unfortunately, it is not possible to constrain the temperature using the single CO line to which we have access. 

The relationship between the Northern and Southern hemispheres of the wind for CO are very similar to those for H~\textsc{i}. Specifically, while the model spectra qualitatively reproduce the observed ones reasonably well, there are clear tidal features produced by M81 that are not captured, and as a result the fit is somewhat more uncertain. The mass outflow rates we infer are consistent within the uncertainties with those found in the North, but the uncertainties  in the South as substantially larger, so that the 68\% confidence interval spans $\approx 1$ dex. 

\subsection{Warm ionised phase}
\label{ch3_sec:wi_ph}

We finally repeat the main steps in the analysis presented in \autoref{ch3_sec:wn_ph} for the warm ionised phase as traced by the H$\alpha$ observations, with the differences in analysis described in \autoref{sssec:Ha_method}. We consider three possible values for the clumping factor,
$c_\rho = 10, 100, 1000$, spanning a range of values measured in individual H \textsc{ii} regions \citep{Kennicutt84}. We summarize the fitting results to the Northern side of the outflow in \autoref{tab:res_fit_halpha}, but only for three highest-weight combinations of driving mechanism, potential and expansion, for each of the three clumping factors. Contrary to what we find for H~\textsc{i} and CO 2$\to$ 1 line, the fitting for H$\alpha$ favors point gravitational potential and clouds that maintain constant solid angle (or intermediate cross sectional area), and we still cannot differentiate strongly between the driving mechanisms. However, the fit parameters still converge into a tiny island of parameter space, with a near-zero orientation angle $\phi$ and reasonable values of opening angle $\theta_{\rm in}$ and $\theta_{\rm out}$, verifying the cone sheath geometry of warm ionised outflow \citep{Martin98}. Most of the fit parameters are similar under the variation of clumping factors, with the exception of the mass outflow. We see the uncertainty of mass flux is about 10 across three clumping factors, from $5$ M$_\odot {\rm yr}^{-1}$ at $c_\rho = 10$, to $1.62$ M$_\odot {\rm yr}^{-1}$ at $c_\rho = 100$, to $0.34$ M$_\odot {\rm yr}^{-1}$ at $c_\rho = 1000$.

\begin{figure}
    \centering
    \includegraphics[width=\columnwidth]{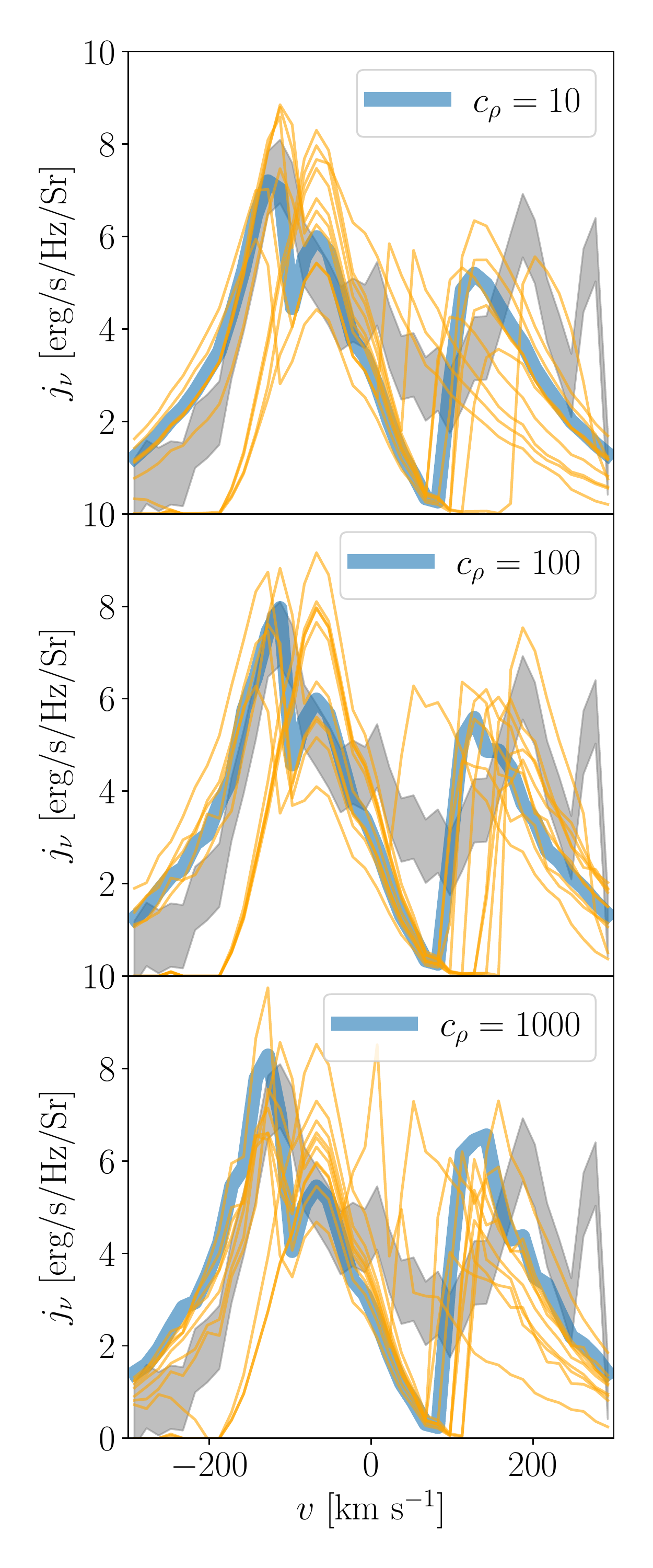}
    \caption{Predicted versus observed H$\alpha$ spectrum for the Northern hemisphere. Each panel shows one choice of clumping factor. Blue lines show the predicted spectra for the set of parameters that gives the largest posterior probability found by our MCMC fit, while orange lines show spectra predicted using the parameters of 10 random walkers at the last iteration of MCMC sampling. For comparison, we also show the observed spectrum with its 1 $\sigma$ errors (grey region).}
    \label{fig:spec_halpha_comp}
\end{figure}

In \autoref{fig:spec_halpha_comp}, we compare predicted and observed spectra, for the ideal, point potential, constant solid angle case, and for three choices of clumping factor. We see that the theoretical spectra give a fairly good fit to the observations, particularly at line wings. All fits qualitatively reproduce the dip near line centre that is the hallmark of cone sheath geometry and an expansion law whereby clouds continually accelerate -- this is feature in the H$\alpha$ data, which is not present in the H~\textsc{i} or CO, is the reason why our model prefers constant solid angle or intermediate expansion for H$\alpha$, and constant area for H~\textsc{i} and CO. However, we also see that there is no clear systematic difference between the spectra predicted for different clumping factors, which limits our model's capacity to distinguish differences in clumping factor. This in turn introduces an order of magnitude uncertainty into the measurement of mass outflow rate. We conclude that the accurate measurement of mass outflow rate for H$\alpha$ data, and by extension for \textit{any} tracer for which the emissivity scales quadratically rather than linearly with the local volume density, is possible only if one has access to independent constraints on this factor.

We also apply the pipeline to the Southern hemisphere. However, we find that the MCMC cannot distinguish a favoured driving mechanism, potential, or expansion law with confidence, and the model-predicted spectra are a poor match to the observations. This is likely because the observed spectrum is severely blue-shifted due to tidal force of M81. For these reasons we do not discuss our fitting results further for the Southern hemisphere.

\begin{table*}
  \centering
  \input table/res_fitting_Halpha.tex

  \caption{Same as \autoref{tab:res_fit_hi}, but showing the results of fits to the H$\alpha$ data. The three blocks in the table show results for three different clumping factors, $c_\rho = 10$, 100, and 1000, as indicated.}
  \label{tab:res_fit_halpha}
\end{table*}

\section{Discussion}
\label{cha:discussion}

Having presented our fits for the multi-phase outflow of M82, we are now in a position to come up with an overall physical picture of the cool wind in M82. We first discuss the mass budget of different phases of the wind in \autoref{ch4_ssec:budget}, and present a broader overview of our model for the wind structure in \autoref{ssec:wind_struct}. We discuss the implications of our results for the origin and the evolution of launched cool gas clouds in \autoref{ch4_ssec:evo_cool_clouds}. We then compare our self-consistent methods of identifying outflow rates and distinguishing winds from fountains to earlier methods in \autoref{ch4_ssec:fountain}, and discuss the implications for the outflow energy and momentum budget in \autoref{ssec:Ebudget} We finally discuss several caveats to our model and their possible effects on our results in \autoref{ch4_sec:caveats}.

\subsection{The M82 wind mass budget}
\label{ch4_ssec:budget}

A primary result of our analysis is that we have, for the first time, obtained tight constraints on the mass outflow rates for the cool phases of the M82 wind. Collating the results from \autoref{tab:res_fit_hi} for the inner extraction region, where we have data in all tracers, and taking a simple average of the 50th, 16th, and 84th percentile values for cases with acceptable Akaike weights (reasonable, since the results are qualitatively identical), we measure mass outflow rates of $\dot{M} = 4.0^{+1.6}_{-1.0}$ M$_\odot$ yr$^{-1}$ and $4.2^{+0.85}_{-0.79}$ M$_\odot$ yr$^{-1}$ in the Northern and Southern sides of the H~\textsc{i} wind, respectively; if we approximate the distributions as Gaussian, and sum the outflow rates in both hemispheres, the 16th to 84th percentile range for the total outflow rate is $\dot{M} = 8.2^{+1.8}_{-1.3}$ M$_\odot$ yr$^{-1}$. The corresponding mass-loading factors in the warm neutral phase, for our fiducial star formation rate $\dot{M}_*=4.1$ M$_\odot$ yr$^{-1}$, are $\eta_{\rm wn} = 1.0^{+0.40}_{-0.25}$ and $1.0^{+0.21}_{-0.20}$ for the Northern and Southern winds separately, and $\eta_{\rm wn} = 2.0^{+0.44}_{-0.31}$ for the sum.

Repeating this exercise for the CO data (\autoref{tab:res_fit_co}), the molecular outflow rates in the Northern and Southern sides separately are $\dot{M} = 2.0^{+1.1}_{-0.65}$ M$_\odot$ yr$^{-1}$ and $0.70^{+4.2}_{-0.13}$ M$_\odot$ yr$^{-1}$, respectively, and the total molecular outflow rate is $\dot{M} = 2.7^{+4.3}_{-0.66}$ M$_\odot$ yr$^{-1}$. The corresponding mass loading factors in the cold molecular phase are $\eta_{\rm cm} = 0.48^{+0.27}_{-0.16}$ and $0.17^{+1.0}_{-0.031}$ for the Northern and Southern sides separately, and $\eta_{\rm cm} = 0.66^{+1.1}_{-0.16}$ for the total. We remind readers that, in addition to our quoted statistical uncertainties, we also have a factor of $\sim 2$ systematic uncertainty arising from the unknown gas temperature in the wind.

We can therefore draw a few high-level conclusions. The total mass outflow rate for the neutral (atomic plus molecular) phases of M82 is $\approx 10$ M$_\odot$ yr$^{-1}$, and the total mass loading factor for these phases $\approx 2-3$. We can be confident in these numbers to better than a factor of two uncertainty. The atomic phase carries moderately more mass than the molecular phase, but both contribute at order unity. We are much less certain about the contribution from the warm ionised phase traced by H$\alpha$, due to the uncertain clumping factor. It could be as much as comparable to the neutral phases, if the clumping factor is small, or it could be an order of magnitude less, if the clumping factor is large. For comparison, \citet{Martin98} estimate a mass outflow rate of $24$ M$_\odot$ yr$^{-1}$ for an assumed H$\alpha$ volume filling factor (equivalent to $1/c_\rho$, where $c_\rho$ is our clumping factor) of $0.1$. This is a factor of several larger than our estimate of $\approx 5$ M$_\odot$ yr$^{-1}$ at $c_\rho = 10$, but given the large differences in assumed geometry, a difference of this size is not particularly surprising. 

\subsection{Structure of the wind}
\label{ssec:wind_struct}

We can also characterise the structure of the wind deduced from our fits more broadly. Since our goal here is qualitative understanding rather than detailed statistical analysis, for clarity we will show only a single result for each CO and H~\textsc{i} component, omitting H$\alpha$ since our fits for it are generally poor, and using only the results from the inner region, where the CO and H~\textsc{i} data overlap. The models we show are those derived from the single highest-likelihood model found by our MCMC for that component, using ideal driving, constant area clouds, and a point-like potential, which is in all cases is one of the models with high Akaike weight. We pause to note that our finding that a point-like potential generally fits the data better than an isothermal potential implies, per our discussion in \autoref{sssec:phys_model}, that either the gas is more reflective of the potential shape than the stars, or that the potential is flattened enough that most of the region that we survey, which begins $0.8$ kpc above the plane, is above the height where the isothermal-Keplerian transition occurs.

\begin{figure}
    \centering
    \includegraphics{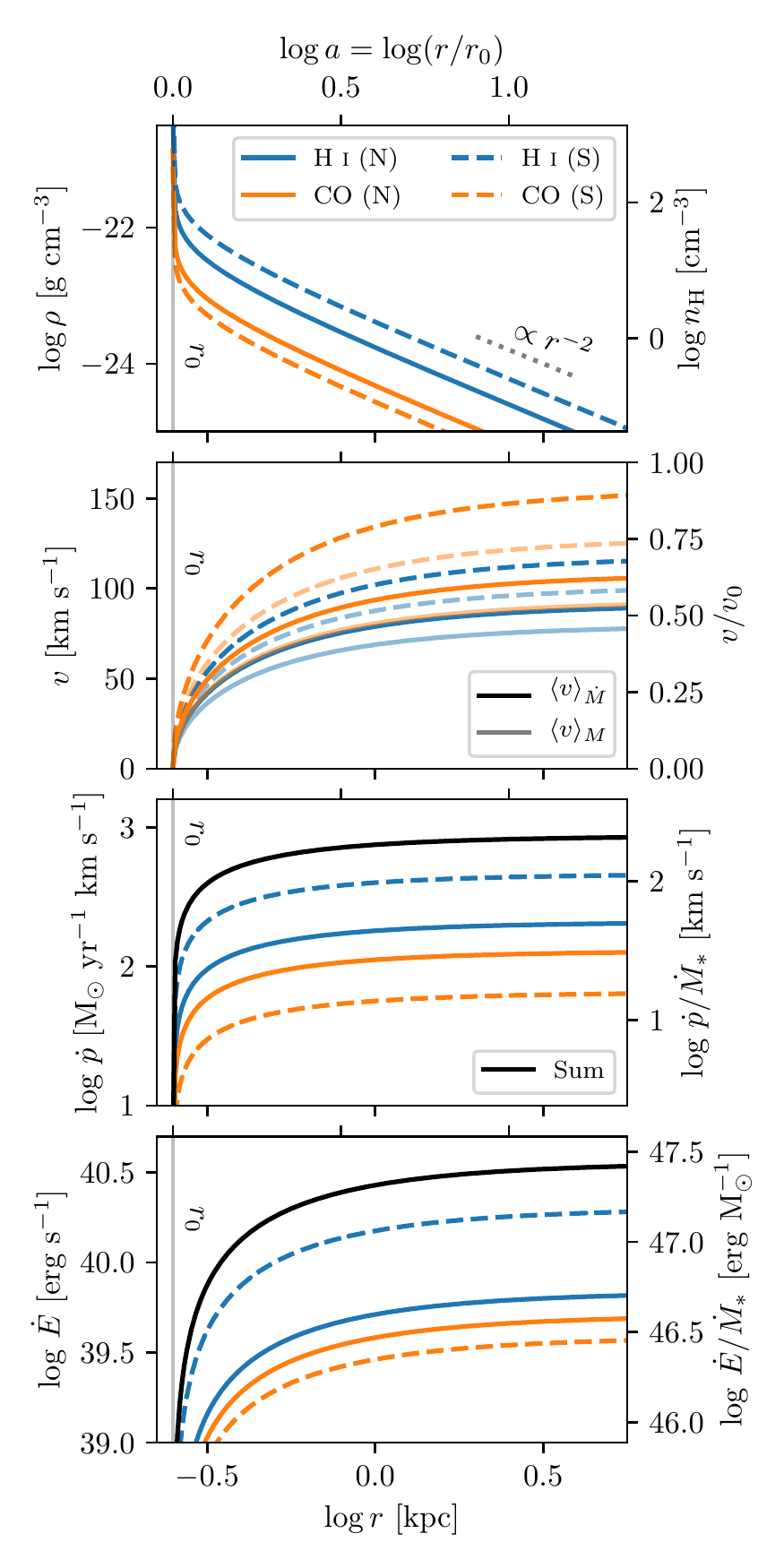}
    \caption{Radial profiles of, from top to bottom, mean density, mean radial velocity, momentum flux, and energy flux carried by the different wind components. In all panels, blue shows H~\textsc{i} and orange shows CO, solid lines show the Northern hemisphere and dashed lines show the Southern hemisphere, and the grey vertical line shows the wind launch radius $r_0 = 250$ pc. For density, the black dotted line shows as $\rho\propto r^{-2}$ scaling for comparison; the left axis is mass density, and the right axis is number density of H nuclei, computed assuming a mean mass of $2.34\times 10^{-24}$ g per H nucleon, as expected for Solar abundances. In the velocity panel, dark lines show the mass flux-weighted mean and light lines show the mass-weighted mean; see \autoref{eq:v_mean}. For the momentum and energy fluxes, the solid black line is the sum of all components; the left axis shows total flux, and the right axis shows flux normalised by star formation rate, so that the quantity shown is the rate per unit mass of stars formed at which stellar feedback must supply momentum and energy to drive the wind to that radius.}
    \label{fig:wind_structure}
\end{figure}

We show the radial profiles of mean density, mean velocity, radial momentum flux, and radial energy flux in \autoref{fig:wind_structure}. We derive these mean values following equations 13 and 17 of \citetalias{Krumholz17}. Specifically, for each component we use the parameters of the best-fitting model derive the mass flux $\dot{M}$, the fraction of area covered by the wind $f_A$, the dispersion $\sigma_x$ of the lognormal surface density distribution $p_M(x)$ (equation 10 of \citetalias{Krumholz17}), the critical logarithmic surface density $x_{\rm crit}$ below which material is ejected, and the mass fraction $\zeta_M$ below this surface density. From these, plus the wind acceleration law $u_a(x)$ for the highest-likelihood model (\autoref{eq:acclaw}), we define the mean density and the momentum and energy fluxes at any specified radius $a = r/r_0$ by
\begin{eqnarray}
    \rho & = & \frac{\dot{M}}{4\pi f_A r_0^2 v_0} \left[\frac{1}{\zeta_M} \int_{-\infty}^{x_{\rm crit}} \frac{p_M}{a^2 u_a} \, dx\right] \\
    \label{eq:pdot}
    \dot{p} & = & \dot{M} v_0 \left[\frac{1}{\zeta_M} \int_{-\infty}^{x_{\rm crit}} u_a p_M \, dx\right] \\
    \label{eq:Edot}
    \dot{E} & = & \frac{1}{2}\dot{M} v_0^2 \left[\frac{1}{\zeta_M} \int_{-\infty}^{x_{\rm crit}} u^2_a p_M \, dx\right].
\end{eqnarray}
Since the differential mass of material launched with surface density $x$ scales as $p_M(x)$, and the differential mass flux as $p_M(x) u_a(x)$, these definitions also immediately permit us to define the mass-weighted and mass flux-weighted mean velocities as
\begin{equation}
\label{eq:v_mean}
    \langle v\rangle_M = \frac{\dot{p}}{\dot{M}} \qquad \langle v\rangle_{\dot{M}} = \sqrt{\frac{2\dot{E}}{\dot{M}}}.
\end{equation}

There are a few features of \autoref{fig:wind_structure} worthy of comment. First, the density and velocity structure asymptote to the values expected for a constant-velocity wind ($v\sim\mbox{constant}$ and $\rho\propto r^{-2}$) only at radii of a few kpc. At smaller radii, the wind is still accelerating, and as a result the density profile is steeper than $r^{-2}$. Second, both the mass-weighted and mass flux-weighted mean velocities are rather small, $\lesssim 150$ km s$^{-1}$, and at no radius does the mean velocity exceed the escape speed from the launch region, $v_0 = 170$ km s$^{-1}$. While there is a tail of material extending to higher velocity (as is required to explain the observed line wings), most of the mass moves much more slowly. The reason it is possible for material moving below the escape speed to nevertheless escape the galaxy is that the gas is not ballistic; instead, the driving mechanism supplies momentum continuously rather than instantaneously, so the net force on gas entrained into the wind is outward rather than inward. Third, this slow acceleration of the wind is reflected in where the wind acquires it energy. The wind acquires $50\%$ of its final, terminal kinetic energy $\gtrsim 0.5$ kpc away from the galaxy, and $25\%$ $\gtrsim 1$ kpc away.

\begin{figure}
\includegraphics[width=\columnwidth]{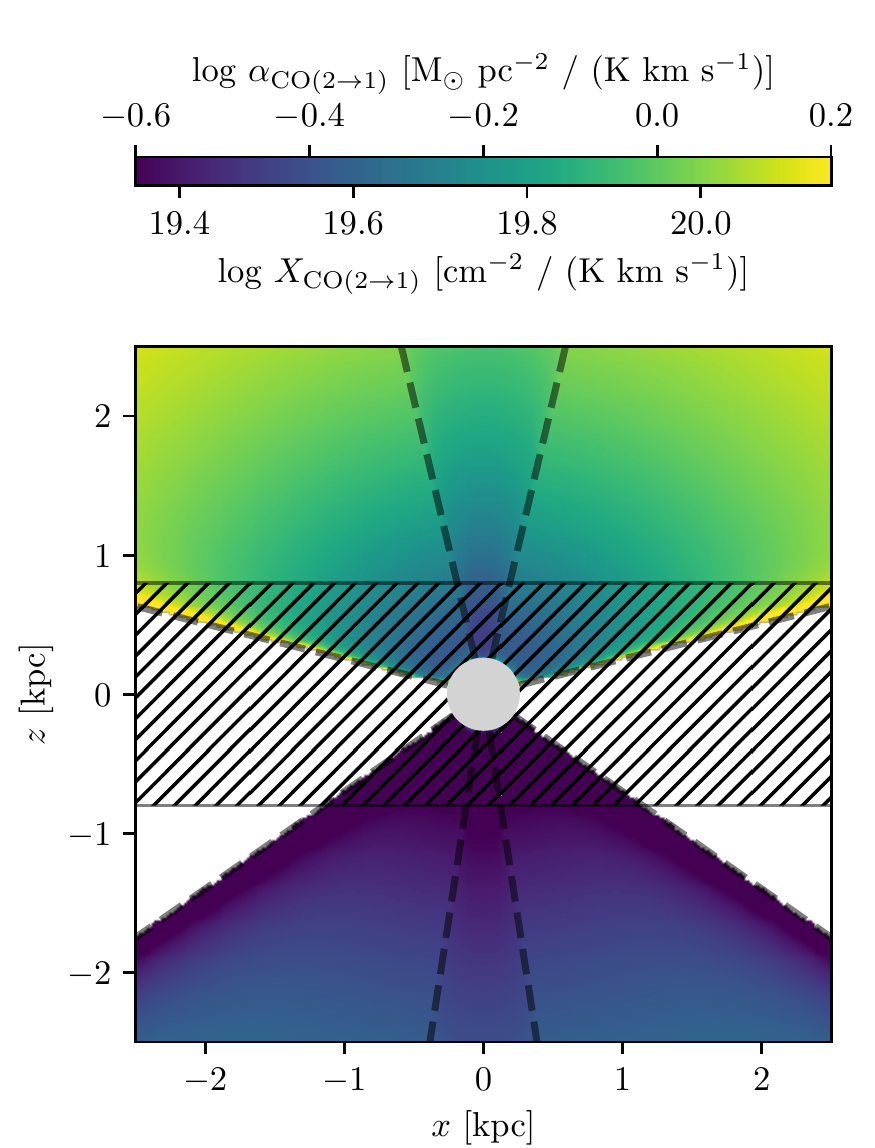}
\caption{
$X_{\mathrm{CO}(2\to 1)}$ or $\alpha_{\mathrm{CO}(2\to 1)}$ as a function of position with the M82 wind; the coordinate system here is the same as that used in \autoref{ch2_fig:pos_spectra}. Dashed lines indicate the inner and outer wind cone edges projected on the sky, and the grey around the origin indicates the wind launch region, where our model is not valid. The hatching marks the region $|z| < 0.8$ kpc that we mask in our analysis due to contamination from disc emission; our model is fit only to data outside this region.
\label{fig:XCO}
}
\end{figure}

We can also use this model to explore the extent to which the CO ``$X$'' factor (or equivalently the $\alpha$ factor) varies across the wind, which is important for interpreting CO observations. Recall that in our model we do not assume a light to mass conversion for CO; instead, we compute it self-consistently from the large-velocity gradient approximation (equation 65 of \citetalias{Krumholz17}). We show the results in \autoref{fig:XCO}. We see that there is a factor of $\sim 2$ difference between the North and South sides of the wind; this is likely a reflection not of a real difference, so much as a reflection of the range of uncertainty in our model; there are other parameter options found by the MCMC with fits that are nearly as good, but where the North-South differences are smaller. For the North side, we generally have $\alpha_{\mathrm{CO}(2\to 1)}\sim 1$ M$_\odot$ pc$^{-2}$ / [K km s$^{-1}$],  comparable to or slightly smaller than the estimate provided by \citet{Leroy15}; or values on the South side are a factor of $\sim 2$ smaller, but, as noted above, this level of disagreement is not surprising given both \citeauthor{Leroy15}'s stated factor of $\approx 2$ systematic uncertainty and the comparably large uncertainty in our analysis that arises from our need to assume a gas temperature.

A more interesting disagreement is the trend of $\alpha_{\mathrm{CO}(2\to 1)}$ with position: \citeauthor{Leroy15} find higher values of $\alpha_{\mathrm{CO}(2\to 1)}$ in the bright parts of the wind than in the faint parts, while we find the opposite. It is unclear which estimate is correct. \citeauthor{Leroy15} derive their conclusions from the dust infrared emission, and their analysis relies on the assumption that the dust in the wind has the same basic emission properties (implying similar temperatures and sizes) as the dust in the disc; this hypothesis may or may not be correct. In our model, where the gas temperature and level populations are assumed to be constant, the trends in $\alpha_{\mathrm{CO}(2\to 1)}$ arise simply from the variation of the line-of-sight velocity gradient across the wind: gas closer to the launch region and to the central outflow axis has larger LOS velocity gradients, and hence smaller $\alpha_{\mathrm{CO}(2\to 1)}$, because those are regions where the wind radial velocity is changing most rapidly, and where the radial direction is most closely aligned with our line of sight. Conversely, regions far from the central axis and the launch point have slowly-varying wind radial velocity, and radial vectors that are mostly in the plane of the sky. These are basic physical effects that it seems must be present; however, it is possible that in reality they are mitigated or outweighed by variations in the gas temperature or excitation state with position, which we do not model.

\subsection{Origin and evolution of cool gas clouds}
\label{ch4_ssec:evo_cool_clouds}

As discussed in \autoref{cha:intro}, it is unclear if the cool gas observed in galactic winds is primarily entrained as cold gas, or if it represents hot gas that has cooled and condensed at some distance from the galactic plane. In M82 we see cool gas spread continuously at locations from immediately above the galactic plane to several kpc off it, which suggest that we are observing in this system is entrainment rather than recondensation. However, this raises the further question of how these cold clouds are able to survive entrainment into the hot wind without disruption.

Our results provide insight into possible solutions, since they suggest that, when launched, cold molecular and warm neutral clouds maintain approximately constant cross-sectional area as they propagate outwards, so their morphologies are well-preserved through the entrainment processes. Previous theoretical studies have not reached consensus on whether, and under what circumstances, this is possible. Purely hydrodynamic simulations from \citet{Hopkins&Elvis10} show that Kelvin-Helmholtz (KH) instabilities stretch cold clouds and increase their cross-sectional area dramatically, while MHD studies from \citep{McCourt15, Banda-Barragn18, Zhang17, Gronnow17} show that the magnetic pressure with $\beta \sim 1$ can suppress mixing \citep{Gentry19}, confining cloud expansion transverse to the direction of magnetic field, and allowing the cloud to survive being entrained. Alternately, expansion can be limited if clouds are accelerated by infrared radiation pressure \citep{Huang20b}, though the modest column density of M82 suggests that infrared radiation pressure is likely unimportant for it \citep{Crocker18a}. Our results therefore favour a magnetic confinement scenario, and hence suggest that M82 is characterised by a strong magnetic field with field lines tracing the trajectory of launched clouds. This scenario is consistent with the recent measurements of magnetic field strength and morphology in M82 \citep{Yoast-Hull13, Jones19, Buckman20,  Lopez-Rodriguez21}, which show that the magnetic field threads from starburst core all the way to intergalactic medium with a strength of $B \sim 300$ $\mu$G and $\beta \sim 0.7$.

\subsection{Outflows versus winds}
\label{ch4_ssec:fountain}
While our fits allow us to quantify with relatively small uncertainty how much gas the winds expel, we are left with another important question: what is the ultimate fate of the gas in the galactic winds in M82? Will it be driven out to the CGM or even further into IGM, or will it be captured by the gravitational potential and fall back as a fountain? Our model fits suggest the former. In the \citetalias{Krumholz17} model, whether a given outflow is a wind or a fountain depends on the indices of the cloud expansion law, $y\propto a^p$, and the gravitational potential, $m\propto a^q$. If $p\geq q$, then the outward force on clouds grows faster than or keeps pace with gravity, so that clouds always move outward and never fall back, while for $p<q$ gravity eventually overwhelms acceleration, resulting in a fountain. Almost all our best fitting models are $p=0$ (constant area) and $q=0$ (point-like potential); the only exceptions are the large-scale H~\textsc{i} in the South, where a model with $p=1$ and $q=0$ (and thus still a wind) is acceptable (Akaike weight $w=0.29$), and H$\alpha$ in the North, where all our acceptable fits again have $p\geq q$ and thus are winds. Therefore our fits generally favour a wind over a fountain, though only by a small amount since we have $p=q$ in most cases.

This conclusion is at odds with those of \citet{Leroy15} and \citet{Martini18}, though we are fitting the same data. It is therefore important to understand why. There are two main features in the data that \citet{Leroy15} and \citet{Martini18} identify as favouring a fountain. The first is that the surface brightness of the CO and H~\textsc{i} emission fall steeply within $\approx 1-2$ kpc of the galactic centre, which \citeauthor{Leroy15} interpret as a steep fall in the mass flux within this region (c.f.~their Figure 18). Our fits also recover this steep fall in surface brightness, but are able do so with a constant mass flux, i.e., a wind rather than a fountain. The underlying reason for this difference in interpretation is a difference in assumptions: \citetalias{Krumholz17} model cool clouds accelerating continuously under the action of some source of momentum deposition, as illustrated in \autoref{fig:wind_structure}, while \citeauthor{Leroy15} assume that cool material is immediately accelerated to its terminal speed near the galactic plane, and after that can only decelerate in response to gravity or drag. The latter assumption imposes strong constraints on the surface brightness profile, because if the velocity can only decrease with radius, then a fixed mass flux implies that density must decrease with radius no more sharply than $\rho\propto r^{-2}$, or, equivalently, that the surface density must decrease with projected radius no more steeply than $r_{\rm proj}^{-1}$. Thus \citeauthor{Leroy15}'s assumption of instantaneous acceleration means that the observed fall in surface brightness with radius, which is steeper than $r_{\rm proj}^{-1}$ near the galactic centre, can \textit{only} be interpreted as a diminution of the mass flux. If one relaxes this assumption, and allows the acceleration to take place over a finite range of radius, then a fit such as that shown in \autoref{fig:wind_structure} becomes possible, whereby the mass flux remains constant and yet the surface density fall more steeply than $r_{\rm proj}^{-1}$ because the wind is still accelerating at small radii.

It is worth emphasising two further points in this regard. First, the \citetalias{Krumholz17} model does not \textit{require} a slowly-accelerating wind; the range in radius over which the wind accelerates is controlled by the parameters of the model. Therefore the fact that the MCMC fit prefers gradual acceleration is significant; gradual acceleration gives a better match to the observed spectra. Second, there is no physical reason to assume that the cool components of winds are instantaneously accelerated. While such an assumption has been common in the literature for reasons of simplicity, there is no physical model for how the cool components of winds are accelerated that produces instantaneous acceleration. Quite the opposite: if the acceleration were instantaneous, the resulting shock would almost certainly dissociate and ionise the cool neutral gas. More gradual acceleration, as in the model we present here, not only fits the data just as well, it is a much more natural physical model.

\begin{figure}
    \centering
    \includegraphics[width=\columnwidth]{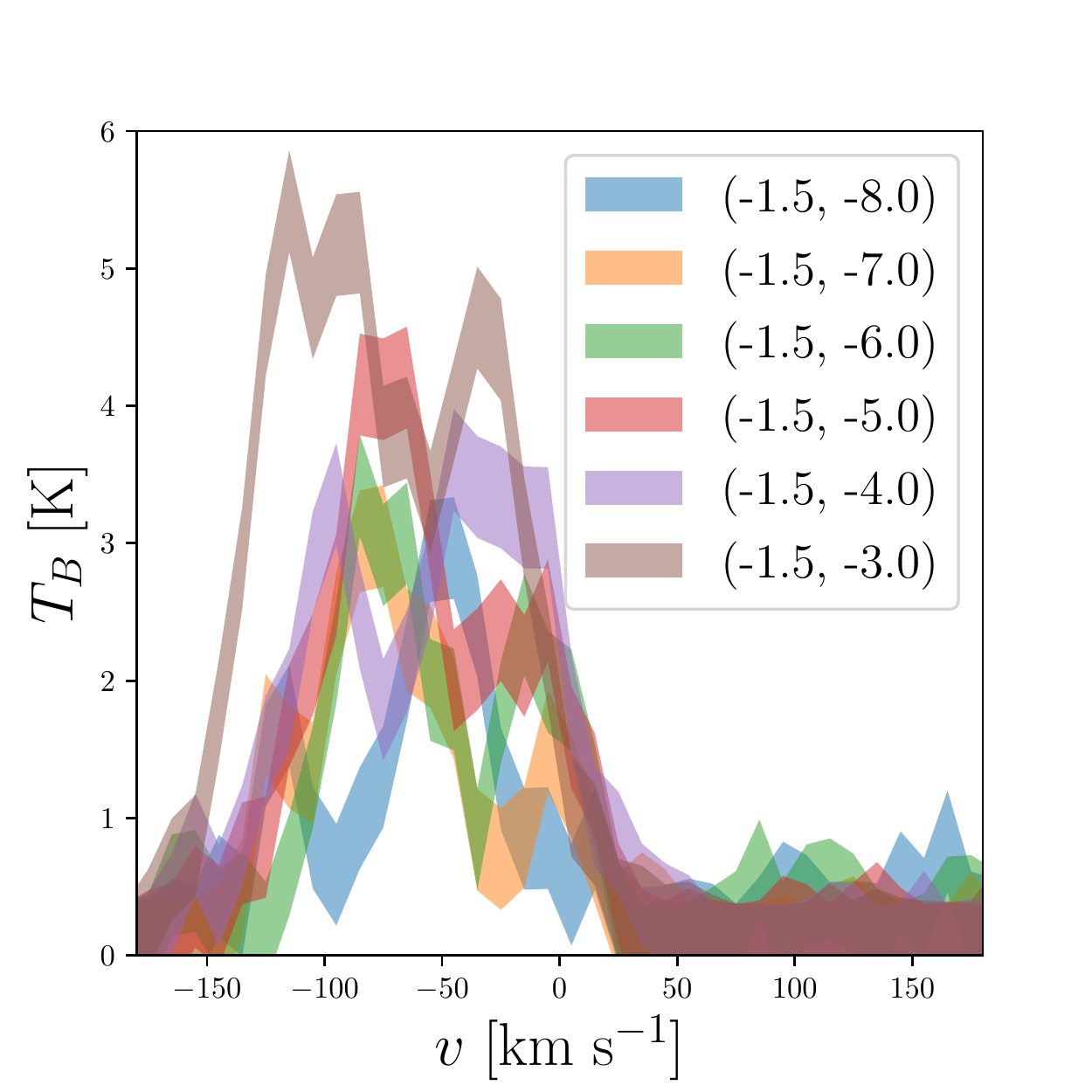}
    \caption{H~\textsc{i} spectra in the Southern hemisphere, at positions from $(x,y) = (-1.5,-3)$ kpc to $(-1.5,-8)$ kpc, as indicated in the legend. Filled bands indicate the uncertainty range.}
    \label{fig:HI_spec_south}
\end{figure}

The other line of evidence that \citet{Martini18} argue favours a fountain rather than an outflow is the position-velocity diagram of the H~\textsc{i} in the Southern hemisphere along the minor axis (c.f.~their Figure 3), which shows a relatively sharp decrease in maximum H~\textsc{i} velocity as one moves away from the galactic centre. They interpret this decrease as evidence that the wind is decelerating. To understand why our fits do not take this decrease as decisive evidence for deceleration, it is helpful to examine the underlying spectra, which we show for the Southern hemisphere along the line $x=-1.5$ kpc\footnote{We show $x=-1.5$ kpc rather than $x=0$, i.e., exactly along the minor axis, because this avoids a region of non-detections along the minor axis. By contrast, at $x=-1.5$ we achieve strong detections of the signal at all vertical distances from $-3$ to $-8$ kpc; c.f.~\autoref{ch2_fig:pos_spectra}.} in \autoref{fig:HI_spec_south}. Examining the figure, it is clear that there is a decrease in emission in the negative velocity wing from $y=-3$ to $-4$ kpc, but decreases at larger radii are much less clear. In particular, it seems possible that the lack of emission beyond $v=-150$ km s$^{-1}$ are larger radii is simply an issue of the signal in the line wing diminishing below the detection limit, rather than an actual decrease in the gas velocity that would be indicative of deceleration. Much the same point can be made with regard to \citeauthor{Martini18}'s Figure 3, and the general effect -- that weak lines systematically underestimate the velocity range of emission because the line wings drop below the detection threshold faster than the line core -- is a well-known bias in atomic and molecular line observations \citep[e.g.,][]{Yuan20a}. Since our model likelihood function properly accounts for the uncertainties on the data (c.f.~\autoref{eq:likelihood}), it will naturally account for this effect, which may be why our fits indicate that winds are acceptable models even for the large-scale H~\textsc{i} emission in the Southern hemisphere.

\subsection{Outflow energy and momentum budget}
\label{ssec:Ebudget}

Our best-fitting model also eases some tension in the energy and momentum budgets for models in which the wind is assumed to flow at constant speed. We use the model proposed by \citet{Leroy15} as an example, but emphasise that the issues we point out are generic to constant-speed wind models. In order for feedback to launch a wind with mass flux $\dot{M}$ at fixed speed $v_w$, it must provide a momentum and kinetic energy per unit mass of stars formed
\begin{eqnarray}
\frac{p}{M_*} & = & \frac{\dot{M}}{\dot{M}_*} v_w \\
\frac{E}{M_*} & = & \frac{\dot{M}}{2 \dot{M}_*} v_w^2,
\end{eqnarray}
where $\dot{M}_*$ is the star formation rate of the stellar population powering the wind. \citeauthor{Leroy15}'s model has $v_w = 450$ km s$^{-1}$, and under their assumption of constant $v_w$, the high surface brightness at small radii requires that the outflow mass flux at launch be $\approx 50$ M$_\odot$ yr$^{-1}$; this only diminishes to $\approx 10$ M$_\odot$ yr$^{-1}$ at $2-3$ kpc above the plane. Inserting these factors into the equations above, we find $p/M_* = 5400$ km s$^{-1}$ and $E/M_* = 2.5\times 10^{49}$ erg M$_\odot^{-1}$.\footnote{These figures are for our fiducial star formation rate $\dot{M}_*=4.1$ M$_\odot$ yr$^{-1}$; adopting \citet{Leroy15}'s preferred star formation rate of $\approx 10$ M$_\odot$ yr$^{-1}$ would halve them, but this would not change the qualitative conclusions we draw.} These figures are difficult to supply: for blast waves driven by single SNe, \citet{Gentry17} find that the terminal momentum and kinetic energy are $\approx 3\times 10^5$ M$_\odot$ km s$^{-1}$ and $\approx 3\times 10^{49}$ erg, while for the most efficient superbubbles produced by clustered SNe (those that suffer the smallest radiative losses), these budgets rise to $\approx 3\times 10^6$ M$_\odot$ km s$^{-1}$ and $\approx 10^{50}$ erg per SN. For a standard IMF that produces approximately one supernova per 100 M$_\odot$ of stars formed, we therefore have firm upper limits of $p/M_* \approx 3\times 10^4$ km s$^{-1}$ and $E/M_* \approx 10^{48}$ erg M$_\odot^{-1}$ for superbubbles, and $3000$ km s$^{-1}$ and $3\times 10^{47}$ erg M$_\odot^{-1}$ for non-clustered SNe. The constant velocity model for the wind fits within the momentum budget for superbubbles, but it exceeds the energy budget by a factor of $\approx 30$; for single SNe, it exceeds the momentum budget by a factor of 2, and the energy budget by a factor of 100.\footnote{Our conclusions about the energy budget are different from those of \citet{Leroy15}, who argue that their model fits within $\approx 30\%$ of the available energy. Partly this is because \citeauthor{Leroy15} adopt a mass flux of $\approx 10$ M$_\odot$ yr$^{-1}$, which their model reaches only $\approx 3$ kpc above the plane, rather than the $\approx 50$ M$_\odot$ yr$^{-1}$ with which it starts. However, a larger issue is that \citeauthor{Leroy15} assume that all of the $\approx 10^{51}$ erg provided by a SN can be converted to kinetic energy of the outflow, and none is lost to radiation. Neither analytic models nor simulations support such an assumption. \citet{Gentry17} find that even in the most energy-efficient superbubbles put $\approx 90\%$ of the initial SN energy into radiation, while for single SNe this rises to $\approx 97\%$. Thus the largest plausible mechanical energy budget is a factor of 10 smaller than the value that \citeauthor{Leroy15} adopt.}

The energy and momentum requirements of a \citetalias{Krumholz17} wind are smaller. This in part because relaxing the assumption of constant speed allows the mass flux at the base to be smaller, and in part because relaxing the assumption that all wind material travels at a single speed allows the high-velocity line wings to be produced by a relatively small mass of low-density material that is accelerated to high speed, while the bulk of the mass moves more slowly. Quantitatively, we can compute the wind energy and momentum requirements simply by evaluating \autoref{eq:pdot} and \autoref{eq:Edot} in the limit $a\to\infty$, in which case $u_x \to \Gamma e^{x}-1$. Doing so using the highest-likelihood values for ideal, point, constant area models, as in \autoref{ssec:wind_struct}, we find that our wind model requires a momentum and kinetic energy per unit mass of stars -- including both hemispheres and both the H~\textsc{i} and CO components -- of $p/M_* = 210$ km s$^{-1}$ and $E/M_* = 2.7\times 10^{47}$ erg M$_\odot^{-1}$, respectively. (Roughly 75\% of this is in the H~\textsc{i} component, and 25\% in the CO.) Thus our model, unlike the constant velocity model, fits comfortably within the available energy and momentum budgets.

A broader point to take from this analysis is that models in which a galactic wind is assumed to be ballistic or to move at constant speed -- both constant with radius, and constant with density -- can be seriously misleading. These assumptions can make the difference between interpreting observations as a wind versus a fountain, and they can lead to order of magnitude errors in the inferred energy budget.

\subsection{Caveats}
\label{ch4_sec:caveats}

We end this section by noting several caveats for our results, as our analytical model inevitably ignores some physical processes. 

The first is that our model assumes that each phase in the wind is composed of a continuous population of clouds with a temperature and chemical composition that is independent of time and distance from the central galaxy. However, it is possible that galactic winds evolve in thermal or chemical properties as the they propagate outward. Depending on the nature of the change, this could either move gas into or out of the cool phases as the outflows propagate outward. The observed weak cooling in the wind of M82 found by \citet{Hoopes03} suggests that cooling of hot phases into the cool phase is unlikely to be a strong effect in this galaxy, but it is unclear if this result applies to outflows in general, or only to M82. Conversely, clouds may dissociate from molecular gas into atomic gas, causing the density of molecular gas to fall and that of atomic gas to rise, or they may ionise, causing mass to disappear in H~\textsc{i} and appear in tracers such as H$\alpha$. Determining how these thermal and chemical processes affect the cool wind requires future observations of the chemical and thermal structure of the wind of M82 and a more accurate description of the thermal and chemical properties of the wind in our model.  

The second limitation of our modelling originates from our ignorance of the larger environment around M82, which may induce complex structures in the real outflows that are not accounted for in our simple model. For example, we have already seen that there is evidence that the torque from M81 induces a tidal stream from M82, and this might alter the structure and distribution of matter in the outflow of M82. This effect is most obvious for the warm neutral outflow traced by H~\textsc{i}, which shows a `northwest spur' and `southeast spur' induced by the drag exerted by ambient environment \citep{Martini18}, This `NW-SE' asymmetry is in contrast with our model prediction, which is symmetric about the minor axis.

\section{Conclusion}
\label{cha:conc}

\subsection{The wind of M82}

Analyses of galactic wind observations have long been limited lack of theoretical tools to extract information from the rich, 3D data sets that we can obtain from line observations of nearby galaxies. In this paper, we make a first step to improve this situation by using a novel semi-analytic wind model to constrain the properties of the multi-phase outflow of the nearby starburst M82. We model the outflow as a continuous population of gas clouds being momentum-driven out of a turbulent galactic disc. The kinematic structure of the wind is then determined by model parameters describing the outflow geometry and mass flux, and physical prescriptions for the outflow driving mechanism, the gravitational potential, and the rates at which cool clouds expand as they flow outward. We use this model to fit position-position-velocity (PPV) data measured in the H~\textsc{i} 21 cm, CO $J=2\to 1$, and H$\alpha$ lines for the wind of M82. Our main conclusions are as follows:

\begin{itemize}
\item Our best-fitting model shows good overall agreement with the observations, both for the sample spectra used for the fits and for the full 2D moment maps, although several areas of tension still remain due to the simplified geometry and physics recipes adopted by our model.
\item The wind has an edge-on orientation and a conical sheath geometry for all phases, consistent with earlier visual analyses. The atomic outflow has a very broad outer opening angle $(\gtrsim 80^\circ)$, reaching nearly to the equator, while the CO is still broad, but somewhat more narrowly confined (outer opening angle $\approx 60^\circ$).
\item The total mass flux carried by the warm neutral phase is $\approx 8$ M$_\odot$ yr$^{-1}$, while that carried by the molecular phase is $\approx 2$ M$_\odot$ yr$^{-1}$. The latter is uncertain at the factor of $\approx 2$ level, while the former is tightly constrained, with only $\approx 25\%$ uncertainty. These fluxes are similar to each other and to the SFR at the order-of-magnitude level, implying a total mass-loading factor of a few. The mass flux is the ionised component is far less certain, due to the quadratic dependence of H$\alpha$ emissivity on the local volume density. We tentatively conclude that the ionised gas mass flux is no more than that in the neutral phases, but are unable to draw any stronger conclusions.
\item We find that the atomic and molecular clouds retain near constant area as the flow outward, rather than expanding to intercept more momentum. This suggests a picture where strong magnetic fields in M82 thread the launched gas clouds and prevent them from expanding and subsequently being destroyed by Kelvin-Helmholtz instability as they move outward along their trajectories.
\item At least over the region covered by the H~\textsc{i} and CO data cubes, the outflow is most consistent with being a wind that will escape the galaxy, rather than a fountain that will fall back. Attempts to deduce whether the outflow is a wind or a fountain that make the simplifying assumption that the outflow is instantaneously accelerated to its final velocity, rather than being allowed to accelerate continuously, likely produce misleading conclusions.
\end{itemize}

\subsection{Future prospects}

A natural question to follow this work is to what extent similar detailed model-fitting can be used in other systems, and for which systems it is likely to be the most fruitful. One obvious point to draw from our analysis is that the cleanest, most unambiguous results -- and, in M82, the best fits -- come from H~\textsc{i} observations. These have the advantage over molecular data that there is no need to adopt a gas temperature or to assume that the molecules are in local thermodynamic equilibrium, and has the advantage over collisionally-excited lines such as H$\alpha$ (or C~\textsc{ii}) that there is no need to worry about a clumping factor. Unfortunately, due to the weakness of the 21 cm line, measurements of H~\textsc{i} emission from winds are scarce. The largest extant sample of extrplanar H~\textsc{i} is the HALOGAS sample of 15 galaxies \citep{Marasco19a}. However, MeerKAT is expected to increase this number dramatically over the next few years \citep{Maddox21a}. These observations are very promising targets for the methods we have deployed here, and application of the \citetalias{Krumholz17} model to them should similarly allow constraints on the neutral gas wind mass flux to better than a factor of two.

Another possibly powerful approach is to use molecules, but reduce the uncertainty about the temperature and excitation state by observing multiple transitions and/or multiple isotopologues. Within galactic discs, measuring spatially-resolved measurements in a sufficiently large set of such lines allows strong constraints on the position-dependent gas temperature and degree of thermal equilibration \citep[e.g.,][]{Sharda21c}; in principle the same methods could be used to derive these quantities for galactic winds, thereby removing our single largest source of systematic uncertainty.

A third approach that we have not explored here, but that can be used directly within the context of the \citetalias{Krumholz17} model, is absorption line measurements. Such measurements also provide a method to constrain outflows that is not dependent on assumed gas temperatures or  clumping factors, and surveys have begun to gather samples of reasonable size \citep[e.g.,][]{Schroetter16a, Schroetter19a}. Here the primary systematic uncertainty is likely to be the abundances of the absorbing species, coupled to the fact that such measurements will establish the outflow rate only in the warm ionised phase, not in other phases. Unfortunately the set of galaxies for which large-scale absorption measurements exist and the set for which atomic and molecular measurements exists is essentially non-overlapping. Thus it is likely to be some time before we can obtain strong constraints on the outflow rates in both the neutral and warm ionised phases for the same galaxy.

\section*{Acknowledgements}

We thank A.~Leroy and P.~Martini for providing us with access to their H~\textsc{i} and CO data. MRK acknowledges support from the Australian Research Council through Future Fellowship award FT180100375. This research was undertaken with the assistance of resources and services from the National Computational Infrastructure (NCI), grant jh2, which is supported by the Australian Government. We also gratefully acknowledge the contribution of Mike Fitzpatrick who wrote a python script to byte swap the echelle IRAF images,  making them readable on modern computers. 

\section*{Data availability}
The source code and data used in this paper are available from \url{https://github.com/yyx319/multiphase-outflow-M82}.



\bibliographystyle{mnras}
\bibliography{refs} 




\appendix

\end{document}

%% file: table/despotic_par.tex
\begin{tabular}{ccc}
\hline\hline
Parameter & Description & Possible range / values \\ \hline
\multicolumn{3}{c}{Continuous parameters} \\
\hline
$\log(\dot{M}/\mathrm{M}_{\odot}\;\mathrm{yr}^{-1})$ & Wind mass flux & $(-1,3)$ \\
$\log\mathcal{M}$ & Mach number in the wind launching region & $(0,4)$ \\
$\phi$ & Inclination of outflow central axis on the plane of the sky & $(-\pi/2,\pi/2)$ \\
$\theta_{\rm in}$ & Inner opening angle of wind cone & $(0,\pi/2)$ \\
$\theta_{\rm out}$ & Outer opening angle of wind cone & $(\theta_{\rm in}, \pi/2)$ \\
$\tau_0$ & Mean optical depth of wind launching region (radiatively-driven winds only) & $(1/\Gamma, 300)$ \\
$u_h$ & Ratio of hot gas speed to escape speed (hot gas-driven winds only) & $(1,50)$ \\
\hline
\multicolumn{3}{c}{Discrete parameters} \\ \hline
$y$ & Scaling of cloud size with radius: constant area, intermediate, constant solid angle & $y=1,a,a^2$ \\
$m$ & Scaling of gravitational potential with radius: point or isothermal potential & $m=1,a$ \\
$f_p$ & Wind driving mechanism: ideal, radiation pressure, hot gas & $f_p = 1, 1-\exp(-e^x\tau_0/y), (1-u/u_h)^2$
\\
\hline\hline
\end{tabular}

%% file: table/res_fitting_HI.tex
{
\begin{center}
\scalebox{1.0}{
    \begin{tabular}{lllrrrrrrrr} \hline
    \multicolumn{11}{c}{H~\textsc{i} 21 cm \citep{Martini18}} \\ \hline
      \multicolumn{3}{c}{Models} & & \multicolumn{7}{c}{Best fit parameters} \\
      Driver & Potential & Expansion & $w$ & $\phi$ & $\theta_{\mathrm{in}}$ & $\theta_{\mathrm{out}}$ & $\log\Dot{M}$ & $\log\mathcal{M}$ & $\tau_0$ & $u_h$ \\ \hline
      \multicolumn{11}{c}{ Northern hemisphere } \\ \hline
     \multirow{7}{*}{Ideal} & \multirow{3}{*}{Point} & Area & \textbf{0.52} & $\mathbf{4.73_{-5.10}^{+4.21}}$ & $\mathbf{30.08_{-2.64}^{+3.54}}$ & $\mathbf{85.96_{-6.15}^{+2.76}}$ & $\mathbf{0.61_{-0.11}^{+0.14}}$ & $\mathbf{\gtrsim 1.57}$ & - & - \\
     &     & Intermediate &   0 & $-37.55_{-2.96}^{+63.91}$ & $38.99_{-8.07}^{+1.97}$ & $69.41_{-13.78}^{+2.55}$ & $1.39_{-0.24}^{+0.04}$ & $\gtrsim 1.92$ &   - & - \\
     &     & Solid angle &   0 & $40.37_{-91.88}^{+16.76}$ & $48.23_{-11.60}^{+4.98}$ & $68.80_{-6.53}^{+4.05}$ & $1.61_{-0.27}^{+0.06}$ & $1.41_{-0.71}^{+0.27}$ &   - & - \\[0.5ex] \hhline{~----------} \noalign{\vspace{0.5ex}}
     & \multirow{3}{*}{Isothermal} & Area &   0 & $3.91_{-6.85}^{+8.71}$ & $11.91_{-3.36}^{+5.96}$ & $75.84_{-8.02}^{+7.70}$ & $1.21_{-0.07}^{+0.05}$ & $\gtrsim 2.84$ &   - & - \\
     &     & Intermediate & 0.02 & $4.16_{-4.22}^{+10.39}$ & $14.58_{-6.12}^{+41.44}$ & $77.84_{-8.63}^{+8.34}$ & $0.71_{-0.25}^{+0.17}$ & - &   - & - \\
     &     & Solid angle &   0 & $-0.52_{-40.44}^{+38.77}$ & $30.53_{-22.24}^{+21.60}$ & $72.22_{-15.41}^{+13.37}$ & $0.36_{-1.71}^{+1.65}$ & - & - & - \\ \hline
     \multirow{7}{*}{Radiation} & \multirow{3}{*}{Point} & Area & \textbf{0.22} & $\mathbf{4.46_{-6.25}^{+6.19}}$ & $\mathbf{28.10_{-18.63}^{+7.04}}$ & $\mathbf{82.68_{-11.67}^{+5.51}}$ & $\mathbf{0.63_{-0.15}^{+0.16}}$ & $\mathbf{\gtrsim 1.50} $ & $\mathbf{76.17_{-23.73}^{+93.99}}$ & - \\
     &     & Intermediate &   0 & $2.84_{-39.36}^{+31.42}$ & $25.85_{-17.48}^{+23.03}$ & $72.75_{-16.06}^{+12.81}$ & $0.44_{-1.34}^{+1.62}$ & - & $64.30_{-38.31}^{+119.05}$ & - \\
     &     & Solid angle &   0 & $-2.52_{-36.65}^{+37.70}$ & $23.80_{-15.23}^{+19.36}$ & $67.20_{-12.52}^{+15.30}$ & $0.63_{-1.99}^{+1.57}$ & - & $81.87_{-53.75}^{+121.21}$ & - \\[0.5ex] \hhline{~----------} \noalign{\vspace{0.5ex}}
     & \multirow{3}{*}{Isothermal} & Area &   0 & $7.27_{-6.56}^{+12.93}$ & $14.53_{-5.24}^{+15.59}$ & $76.91_{-9.60}^{+8.36}$ & $1.22_{-0.08}^{+0.07}$ & $3.08_{-1.23}^{+0.92}$ & $72.16_{-29.63}^{+107.31}$ & - \\
     &     & Intermediate & 0.01 & $3.96_{-3.94}^{+8.63}$ & $27.75_{-17.74}^{+6.47}$ & $80.07_{-8.25}^{+6.65}$ & $0.67_{-0.17}^{+0.14}$ & $1.41_{-0.99}^{+1.11}$ & $102.83_{-42.36}^{+83.02}$ & - \\
     &     & Solid angle & 0 & $16.17_{-58.68}^{+31.19}$ & $28.27_{-12.49}^{+14.78}$ & $63.98_{-4.99}^{+5.55}$ & $0.88_{-0.50}^{+0.62}$ & $2.03_{-0.53}^{+0.34}$ & $41.17_{-13.49}^{+14.20}$ & - \\ \hline
     \multirow{7}{*}{Hot gas} & \multirow{3}{*}{Point} & Area & \textbf{0.23} & $\mathbf{3.10_{-5.37}^{+5.11}}$ & $\mathbf{54.51_{-24.74}^{+7.24}}$ & $\mathbf{78.17_{-6.32}^{+6.53}}$ & $\mathbf{0.57_{-0.13}^{+0.14}}$ & $\mathbf{\gtrsim 1.38}$ & - & $\mathbf{19.84_{-7.85}^{+7.47}}$ \\
     &     & Intermediate &   0 & $-36.08_{-3.52}^{+78.08}$ & $38.63_{-4.44}^{+2.15}$ & $69.41_{-4.15}^{+2.93}$ & $1.33_{-0.19}^{+0.06}$ & $\gtrsim 1.62$ &   - & $15.64_{-4.15}^{+7.47}$ \\
     &     & Solid angle &   0 & $-35.48_{-11.73}^{+87.18}$ & $42.33_{-12.97}^{+7.20}$ & $69.33_{-6.51}^{+4.87}$ & $1.43_{-0.13}^{+0.09}$ & $\lesssim 0.96$  &   - & $15.92_{-4.49}^{+8.27}$ \\[0.5ex] \hhline{~----------} \noalign{\vspace{0.5ex}}
     & \multirow{3}{*}{Isothermal} & Area &   0 & $2.96_{-6.73}^{+5.54}$ & $14.64_{-4.76}^{+15.43}$ & $70.28_{-8.67}^{+12.73}$ & $1.16_{-0.09}^{+0.10}$ & $\gtrsim 2.93$ &   - & $13.18_{-3.80}^{+8.40}$ \\
     &     & Intermediate &   0 & $2.80_{-3.96}^{+4.08}$ & $29.81_{-10.14}^{+4.41}$ & $79.38_{-9.57}^{+6.65}$ & $0.70_{-0.14}^{+0.13}$ & $\gtrsim 1.57$ &   - & $24.63_{-11.85}^{+4.05}$ \\
     &     & Solid angle & 0 & $-40.03_{-10.19}^{+88.70}$ & $42.71_{-9.32}^{+8.82}$ & $69.55_{-6.66}^{+6.15}$ & $1.50_{-0.18}^{+0.10}$ & $1.28_{-0.94}^{+0.55}$ & - & $14.53_{-4.01}^{+7.57}$ \\ \hline
     \multicolumn{11}{c}{ Southern hemisphere } \\ \hline
     Ideal & Point & Area & 0.85 & $6.09_{-36.60}^{+5.77}$ & $42.86_{-32.34}^{+7.57}$ & $65.26_{-6.51}^{+12.53}$ & $0.62_{-0.10}^{+0.08}$ & $1.65_{-0.46}^{+0.50}$ & - & -   \\ 
     Radiation & Point & Area & 0.15 & $7.74_{-42.15}^{+4.60}$ & $32.23_{-25.81}^{+11.20}$ & $68.33_{-7.67}^{+9.41}$ & $0.63_{-0.08}^{+0.08}$ & $1.71_{-0.39}^{+0.46}$ & $55.88_{-20.82}^{+18.29}$ & - \\
     \hline
    \hline 
    \multicolumn{11}{c}{ Large scale Northern hemisphere } \\ \hline
     Ideal &  Point &  Area & $0.65$ & $-0.25_{-14.60}^{+22.61} $ & $22.35_{-14.93}^{+36.49} $ & $84.40_{-9.37}^{+4.27}$ & $0.79_{-0.15}^{+0.11}$ & $1.97_{-1.08}^{+0.70}$ &   - & - \\
     Ideal & Isothermal & Area & 0.01 & $2.95_{-11.66}^{+11.98}$ & $17.08_{-9.38}^{+8.06}$ & $67.72_{-8.27}^{+8.03}$ & $1.63_{-0.13}^{+0.10}$ & $1.73_{-0.37}^{+0.47}$ &   - & - \\
     Ideal & Isothermal   & Intermediate & 0.04 & $3.88_{-18.59}^{+22.68}$ & $18.81_{-12.01}^{+9.02}$ & $77.50_{-13.54}^{+8.64}$ & $0.99_{-0.19}^{+0.18}$ & $1.19_{-0.70}^{+1.02}$ &   - & - \\
     Radiation & Point & Area & 0.20 & $3.98_{-17.91}^{+27.68}$ & $21.33_{-13.66}^{+27.89}$ & $82.69_{-8.62}^{+5.53}$ & $0.79_{-0.14}^{+0.11}$ & $2.06_{-0.99}^{+0.66}$ & $78.76_{-39.33}^{+108.75}$ & - \\
     Hot gas & Point & Area & 0.08 & $-5.67_{-6.41}^{+8.36}$ & $55.08_{-7.58}^{+6.73}$ & $77.79_{-6.83}^{+6.27}$ & $0.77_{-0.13}^{+0.12}$ & $1.70_{-0.55}^{+0.72}$ &   - & $21.25_{-7.67}^{+6.48}$ \\
     \hline
    \multicolumn{11}{c}{ Large scale Southern hemisphere } \\ \hline
     Ideal & Point & Area & $0.57$ & $-11.21_{-5.54}^{+34.96}$ & $40.95_{-24.80}^{+20.50}$ & $81.43_{-22.75}^{+6.79}$ & $0.78_{-0.17}^{+0.16}$ & $1.47_{-0.52}^{+0.72}$ &   - & - \\
     Ideal & Point & Intermediate & 0.29 & $-12.71_{-22.05}^{+13.89}$ & $39.85_{-10.25}^{+4.45}$ & $75.66_{-19.13}^{+11.86}$ & $0.90_{-0.14}^{+0.19}$ & $0.41_{-0.32}^{+0.86}$ &   - & - \\
     Radiation & Point & Area & 0.13 & $-10.66_{-7.39}^{+17.36}$ & $40.01_{-8.12}^{+2.81}$ & $80.41_{-20.75}^{+7.72}$ & $0.79_{-0.16}^{+0.17}$ & $1.68_{-0.58}^{+0.74}$ & $67.83_{-27.90}^{+93.86}$ & - \\
     \hline
    \end{tabular}}
\end{center}
}

%% file: table/res_fitting_CO.tex
{
\begin{center}
\scalebox{1.}{
    \begin{tabular}{lllrrrrrrrr} \hline
    \multicolumn{11}{c}{CO $J=2\to 1$ \citep{Leroy15} } \\ \hline
      \multicolumn{3}{c}{Models} &  & \multicolumn{7}{c}{Best fit parameters} \\
      Driver & Potential & Expansion & $w$ & $\phi$ & $\theta_{\mathrm{in}}$ & $\theta_{\mathrm{out}}$ & $\log\Dot{M}$ & $\log\mathcal{M}$ & $\tau_0$ & $u_h$ \\ \hline
      \multicolumn{10}{c}{ Northern hemisphere } \\ \hline
 Ideal & Point & Area & 0.54 & $-3.16_{-12.22}^{+9.57}$ & $11.34_{-7.85}^{+11.36}$ & $80.87_{-10.92}^{+6.30}$ & $0.30_{-0.17}^{+0.19}$ & $0.83_{-0.30}^{+0.22}$ &   - & - \\
     Radiation & Point & Area & 0.24 & $-2.50_{-12.06}^{+11.88}$ & $11.88_{-8.33}^{+11.80}$ & $77.52_{-9.92}^{+8.40}$ & $0.29_{-0.17}^{+0.22}$ & $0.84_{-0.34}^{+0.29}$ & $78.08_{-29.00}^{+99.18}$ & - \\
 Hot Gas & Point & Area & 0.22 & $-4.59_{-9.61}^{+8.73}$ & $14.63_{-9.98}^{+9.93}$ & $80.28_{-9.17}^{+6.84}$ & $0.36_{-0.20}^{+0.15}$ & $0.74_{-0.25}^{+0.28}$ &   - & $14.66_{-5.34}^{+10.09}$ \\
 \hline
     \multicolumn{11}{c}{ Southern hemisphere } \\ \hline
 Ideal & Point & Area & 0.01 & $11.57_{-22.24}^{+18.33}$ & $26.63_{-6.76}^{+7.99}$ & $54.60_{-3.26}^{+12.44}$ & $-0.00_{-0.11}^{+0.83}$ & $1.49_{-1.40}^{+0.21}$ &   - & - \\
 Radiation & Point & Area & 0.04 & $17.97_{-28.48}^{+16.09}$ & $15.38_{-7.12}^{+11.09}$ & $55.05_{-4.06}^{+11.19}$ & $-0.13_{-0.25}^{+0.74}$ & $1.71_{-1.15}^{+0.33}$ & $61.89_{-21.75}^{+35.19}$ & - \\
 Hot gas & Point & Area & 0.95 & $18.96_{-39.51}^{+3.39}$ & $10.72_{-6.92}^{+28.25}$ & $56.22_{-3.29}^{+19.68}$ & $-0.34_{-0.10}^{+0.98}$ & $2.07_{-1.43}^{+0.26}$ &   - & $8.81_{-4.61}^{+4.11}$ \\
    \hline
    \end{tabular}}
\end{center}
}

%% file: table/res_fitting_Halpha.tex
{
\begin{center}
\scalebox{1.0}{
    \begin{tabular}{lllrrrrrrrrr} \hline
    \multicolumn{12}{c}{H$\alpha$ \citep{Martin98} } \\ \hline
      \multicolumn{3}{c}{Models} & & \multicolumn{8}{c}{Best fit parameters} \\
      Driver & Potential & Expansion & $w$ & $\phi$ & $\theta_{\mathrm{in}}$ & $\theta_{\mathrm{out}}$ & $\log\Dot{M}$ & $A$ & $\log \mathcal{M}$ & $\tau_0$ & $u_h$ \\ \hline
      \multicolumn{12}{c}{ North } \\ \hline
    \multicolumn{12}{c}{ $c_\rho = 10$ } \\ \hline
     Ideal & Point & Intermediate & 0.30 & $-0.68_{-2.15}^{+1.52}$ & $40.88_{-5.59}^{+2.98}$ & $77.50_{-8.45}^{+8.41}$ & $0.16_{-0.21}^{+0.24}$ & $0.37_{-0.06}^{+0.08}$ & $1.80_{-0.52}^{+0.70}$ & - & - \\
      Ideal & Point & Solid angle & 0.49 & $-10.95_{-14.09}^{+10.95}$ & $29.66_{-10.57}^{+15.14}$ & $68.65_{-12.77}^{+13.83}$ & $0.73_{-0.24}^{+0.17}$ & $0.46_{-0.07}^{+0.06}$ & $1.98_{-0.59}^{+0.59}$ &   - & - \\
      Radiation & Point & Solid angle & 0.21 & $-1.59_{-10.53}^{+1.84}$ & $28.31_{-3.21}^{+3.69}$ & $79.09_{-11.57}^{+7.90}$ & $0.67_{-0.43}^{+0.23}$ & $0.38_{-0.06}^{+0.09}$ & $1.41_{-0.87}^{+1.00}$ & $74.07_{-13.41}^{+35.31}$ & - \\ \hline
    \multicolumn{12}{c}{ $c_\rho = 100$ } \\ \hline
     Ideal & Point & Intermediate & 0.32 & $-0.61_{-3.42}^{+2.77}$ & $40.30_{-12.68}^{+3.69}$ & $77.08_{-10.28}^{+9.44}$ & $-0.37_{-0.30}^{+0.22}$ & $0.42_{-0.08}^{+0.08}$ & $2.51_{-0.73}^{+0.35}$ &   - & - \\
     Ideal & Point & Solid angle & 0.53 & $-5.94_{-16.22}^{+5.91}$ & $23.64_{-4.23}^{+18.25}$ & $75.54_{-10.84}^{+9.69}$ & $0.21_{-0.34}^{+0.21}$ & $0.44_{-0.09}^{+0.06}$ & $1.87_{-0.98}^{+0.80}$ &   - & - \\
     Radiation & Isothermal & Solid angle & 0.15 & $-0.93_{-8.70}^{+2.11}$ & $28.90_{-5.00}^{+4.00}$ & $80.46_{-10.28}^{+7.13}$ & $-0.47_{-0.33}^{+0.39}$ & $0.37_{-0.08}^{+0.06}$ & $0.50_{-0.28}^{+0.56}$ & $76.57_{-13.72}^{+62.42}$ & - \\
     \hline
    \multicolumn{12}{c}{ $c_\rho = 1000$ } \\ \hline
    Ideal & Point & Intermediate & 0.37 & $-0.93_{-5.02}^{+2.32}$ & $41.32_{-7.62}^{+3.32}$ & $76.16_{-8.74}^{+9.15}$ & $-1.04_{-0.34}^{+0.30}$ & $0.40_{-0.07}^{+0.10}$ & $2.34_{-0.65}^{+0.46}$ &   - & - \\
   Ideal & Point & Solid & 0.53 & $-2.55_{-18.45}^{+3.29}$ & $22.26_{-3.58}^{+22.69}$ & $76.73_{-9.47}^{+7.77}$ & $-0.47_{-0.36}^{+0.28}$ & $0.42_{-0.07}^{+0.08}$ & $1.74_{-0.56}^{+0.64}$ &   - & - \\
    Radiation & Isothermal & Solid & 0.10 & $-1.06_{-13.82}^{+2.96}$ & $31.48_{-6.28}^{+6.21}$ & $78.77_{-12.24}^{+7.80}$ & $-1.46_{-0.42}^{+0.70}$ & $0.38_{-0.08}^{+0.10}$ & $0.38_{-0.23}^{+0.66}$ & $74.30_{-19.24}^{+76.51}$ & - \\ \hline
    \end{tabular}}
\end{center}
}